\documentclass[useAMS,usenatbib]{mn2e}

\usepackage{graphicx}
\usepackage{subfigure}
\usepackage{amsmath}
\usepackage{amssymb}
\DeclareGraphicsExtensions{.pdf, .ps}

\def\gsim{\;\rlap{\lower 2.5pt
 \hbox{$\sim$}}\raise 1.5pt\hbox{$>$}\;}
\def\lsim{\;\rlap{\lower 2.5pt
   \hbox{$\sim$}}\raise 1.5pt\hbox{$<$}\;}

\title[Feedback Heating by Cosmic Rays]{Feedback Heating by Cosmic Rays in Clusters of Galaxies} 
\author[F. Guo and S. P. Oh]{Fulai Guo\thanks{E-mail: johnnie@physics.ucsb.edu} and S. Peng Oh\thanks{E-mail: peng@physics.ucsb.edu} \\
Department of Physics; University of California; Santa Barbara, CA 93106
}

\begin{document}
\bibliographystyle{mn2e}

%\date{      }

%\pagerange{\pageref{firstpage}--\pageref{lastpage}} \pubyear{0000}

\pagerange{000--000} \pubyear{0000}
\maketitle

\label{firstpage}

\begin{abstract}
Recent observations show that the cooling flows in the central regions of galaxy clusters are highly suppressed. Observed AGN-induced cavities/bubbles are a leading candidate for suppressing cooling, usually via some form of mechanical heating. At the same time, observed X-ray cavities and synchrotron emission point toward a significant non-thermal particle population. Previous studies have focused on the dynamical effects of cosmic-ray pressure support, but none have built successful models in which cosmic-ray heating is significant. Here we investigate a new model of AGN heating, in which the intracluster medium is efficiently heated by cosmic-rays, which are injected into the ICM through diffusion or the shredding of the bubbles by Rayleigh-Taylor or Kelvin-Helmholtz instabilities. We include thermal conduction as well. Using numerical simulations, we show that the cooling catastrophe is efficiently suppressed. The cluster quickly relaxes to a quasi-equilibrium state with a highly reduced accretion rate and temperature and density profiles which match observations. Unlike the conduction-only case, no fine-tuning of the Spitzer conduction suppression factor $f$ is needed. The cosmic ray pressure, $P_{c}/P_{g} \lsim 0.1$ and $\nabla P_{c} \lsim 0.1 \rho g$, is well within observational bounds. Cosmic ray heating is a very attractive alternative to mechanical heating, and may become particularly compelling if {\it GLAST} detects the $\gamma$-ray signature of cosmic-rays in clusters.
\end{abstract}


\section{Introduction}

Clusters of galaxies contain a large amount of hot diffuse gas which emits prolifically in thermal X-rays.
The X-ray surface brightness of many galaxy clusters is strongly peaked in the central regions, where the cooling time is much shorter than the Hubble time. In the absence of any heating sources, the radiative cooling due to this emission will induce a subsonic inflow of gas in order to maintain pressure equilibrium, leading to substantial dropout of cold gas in the inner regions of rich clusters. The mass deposition rates at the cooling radius, where the cooling time equals to the age of the cluster, were estimated to be as much as several hundred M$_{\sun}$ yr$^{-1}$ in some clusters \citep[see][for a review]{1994ARA&A..32..277F}. However, recent high-resolution \textit{Chandra} and \textit{XMM-Newton} observations show that while the temperature is declining toward the center, there is  a remarkable lack of emission lines from the gas at temperature below about $\sim 1/3$ of the ambient cluster temperature  (e.g., \citet{2001A&A...365L.104P, 2003ApJ...590..207P, 2001A&A...365L..87T}; for a review see \citet{2006PhR...427....1P}). Moreover, the spectroscopically determined mass deposition rates are about $10$ times smaller than the classic values estimated from X-ray luminosity within the cooling regions \citep{2004MNRAS.347.1130V}, and thus limited to at most $\sim {\rm few} \, \times \, 10 \, {\rm M_{\odot} \, yr^{-1}}$.  
These discrepancies suggest that mass dropout is prevented, or significantly reduced, by heating sources. Alternatively, the gas cools with line emission highly suppressed by mixing, inhomogeneous metallicity distributions, differential absorption, or photoionization \citep{2001MNRAS.321L..20F, 2002MNRAS.332L..50F, 2003MNRAS.338..824M, 2004MNRAS.353..468O}, though a fully successful model which explains the observations on such grounds is still lacking. 

Many mechanisms for heating the intracluster gas have been put forth recently, including transport of heat from the hot outer regions of the cluster by thermal conduction \citep[e.g.][]{2003ApJ...582..162Z, 2004MNRAS.347.1130V} or turbulent mixing \citep{2003ApJ...596L.139K, 2005ApJ...622..205D} and heating by outflows, bubbles or sound waves from a central active galactic nuclei \citep[e.g.,][]{2001ApJ...554..261C,2002Natur.418..301B, 2002ApJ...581..223R, 2004ApJ...611..158R}. 
Recent theoretical and numerical work \citep[e.g.][]{2001ApJ...562L.129N, 2003ApJ...589L..77C} has shown that a turbulent magnetic field may not be as efficient in suppressing thermal conduction as previously thought. In particular, \citet{2001ApJ...562L.129N} showed that the effective thermal conductivity $\kappa$ in a turbulent MHD medium is a substantial fraction ($\sim 1/5$) of the classical Spitzer value $\kappa_{\text{Sp}}$ if magnetic turbulence extends over at least two decades in scale. However, \citet{2004MNRAS.347.1130V} found that thermal conduction alone is usually insufficient to heat the gas in the inner parts of hot clusters and most regions of cool clusters. Moreover, balancing cooling with thermal conduction alone is generally globally unstable \citep[e.g.][]{2003ApJ...596..889K} and, as we shall see in \S~\ref{section:condsim}, requires fine-tuning. 

On the other hand, heating from a central AGN provides a self-regulating feedback mechanism, which may play a key role in halting a cooling catastrophe in the intra-cluster medium (ICM). Here the cooling flow triggers AGN activity and heating, which in turn suppresses the cooling catastrophe. The flow thus automatically adjusts itself to a low value of the accretion rate, depending mainly on the feedback coefficient ($\epsilon$ in eq. [\ref{lbubble}]), so there is no fine-tuning problem. Recent observations suggest that active galactic nuclei (AGNs) at the centers of the clusters may interact with and substantially heat the ICM. Radio-loud activity is very common at the centers of cool core clusters \citep{1990AJ.....99...14B}. Recent high-resolution X-ray observations have also revealed cavities or bubbles having sizes of a few kiloparsecs in many galaxy clusters \citep[e.g.][]{2004ApJ...607..800B}. 

Such AGN feedback heating models have received a good deal of attention in recent years. Numerical simulations of AGN bubbles (and jets, in some cases) in the ICM have been performed by a number of authors \citep[e.g.][]{2002Natur.418..301B, 2004ApJ...611..158R, 2006ApJ...645...83V, 2005MNRAS.357..242R}. These simulations usually focus on the heating of the ICM by the $pdV$ work of the expanding bubbles, viscous dissipation of emitted sound waves or mixing of the hot bubble plasma with the ICM. Most simulations assume that these X-ray cavities are filled with low-density gas at very high temperature (e.g., $\sim 100$ keV). However, radio synchrotron and inverse Compton emission has been observed from many cavities, suggesting a significant  non-thermal component, such as cosmic rays and magnetic fields, in these cavities. Cosmic rays (CRs) can indeed be injected at the tip of a radio jet, which moves supersonically in the ICM at its initial momentum-driven phase and forms radio cocoons or bubbles at its later stage. A substantial amount of cosmic rays may then escape from these buoyantly rising bubbles \citep[e.g.][]{2003A&A...399..409E} and heat the ICM. X-ray observations \citep[e.g.][]{2006MNRAS.366..417F} show that some bubbles do remain stable even far from cluster centers, but a significant fraction of the bubble population could be shredded or disrupted as they rise through the ICM. Indeed, numerical simulations show that the surfaces of buoyant bubbles are highly susceptible to disruption by Rayleigh-Taylor and Kelvin-Helmholz instabilities \citep[e.g.][]{2002Natur.418..301B}, unless viscosity or magnetic fields are invoked to suppress these instabilities \citep[e.g.][]{2005MNRAS.357..242R,2007astro.ph..3801R}. Hence, it is of great interest to study the consequences---particularly the heating effect on the ICM---of cosmic rays leaked or disrupted from radio cocoons or bubbles. 

Cosmic rays may also be produced by other processes near the central AGN. Structure formation shocks, merger shocks and supernovae may also inject cosmic rays into the ICM \citep[e.g.][]{1996SSRv...75..279V, 1997ApJ...487..529B}. Direct evidence for the presence of an extensive population of nonthermal particles in galaxy clusters comes from the observation of diffuse radio synchrotron emission (e.g., radio haloes, mini-haloes and relics) in many massive clusters \citep[e.g.][]{2001MNRAS.320..365B, 2001ApJ...557..560P, 2004A&A...413...17P, 2005AdSpR..36..729F}; recent {\it Chandra} and XMM observations also show evidence for a significant nonthermal particle population within the ICM \citep{2005MNRAS.360..133S,2007arXiv0704.0475W,2007arXiv0705.2712S}. Cosmic rays have also been inferred from the excess abundance of $^{6}$Li in metal-poor halo stars, since $^{6}$Li could be produced in spallation reactions by cosmic rays \citep{2006MNRAS.366L..35N}. 

The cosmic-ray heating of the ICM has been studied by several authors \citep{1988ApJ...330..609B, 1991ApJ...377..392L, 1995ApJ...442...91R, 2004A&A...413..441C,jubelgas06,pfrommer06}. While many studies have found that cosmic rays could in principle be dynamically important, none have constructed successful models in which CR heating prevents a cooling catastrophe. In the steady-state model of \citet{1988ApJ...330..609B}, a significant cooling flow ($\sim 300 {\rm M_{\odot} \, yr^{-1}}$) developed. \citet{1991ApJ...377..392L} proposed a hydrostatic model where radiative cooling is fully balanced by hydromagnetic-wave-mediated cosmic-ray heating and thermal conduction. Their model could not fit the observational gas density profile; in particular, they found that the intra-cluster medium would quickly become CR pressure dominated at a level inconsistent with observations, long before heating effects become significant. \citet{1995ApJ...442...91R} estimated the CR heating rate from Coulomb collisions alone, which they argued could be significant, but did not construct a specific model for the intracluster gas, which could be compared against observations. \citet{2004A&A...413..441C} do construct semi-analytic models of cluster density and temperature profiles, which differ significantly from ours: they only consider Coulomb and hadronic heating; the ICM is not in thermal equilibrium but evolves strongly as a function of time\footnote{This may be hard to reconcile with observational evidence that the temperature profiles of the intracluster gas are well described by an approximately universal mathematical function across a range in redshift \citep{2001MNRAS.328L..37A}.}; they do not solve for hydrostatic equilibrium, and thus their input density profiles $n_{e}(r)$ do not evolve as the temperature profile $T(r)$ evolves. More recently, \citet{jubelgas06} and \citet{pfrommer06} have run a suite of high-resolution 3D numerical simulations analyzing the role of cosmic rays in clusters; these simulations represent the most comprehensive and careful treatment of this problem to date. These authors find that cosmic ray heating cannot stem a cooling flow; in particular they find that the increased compressibility (due to the softer adiabatic index of CRs) can lead to enhanced cooling. However, as they note, this may be due to the relatively modest production of CRs in their self-consistent treatment, where cooling gas gives rise to star formation and hence supernovae (the source of CRs in their model). Moreover, they only consider cosmic ray heating through Coulomb interactions with the ICM, which is much less than the hydromagnetic-wave mediated cosmic-ray heating in our models. The inclusion of AGN-mediated CR production and hydromagnetic wave mediated cosmic-ray heating could significantly alter this picture. 

In contrast to previous work, we find that it is indeed possible to construct models in which
cosmic-ray heating strongly suppresses the cooling flow. Although our results are generic, we choose the typical cool core cluster Abell 2199 as our fiducial model. By using 1D numerical hydrodynamic simulations, we demonstrate that, starting from a state far from thermal equilibrium (isothermal temperature profile), the cluster will relax to a stable steady quasi-equilibrium state, in which the accretion rate is highly reduced and the temperature and density profiles are consistent with observations. We take the relevant cosmic ray physics into account: various CR energy loss mechanisms, heating of the ICM by cosmic rays, cosmic ray pressure support, and cosmic ray transport (advection and diffusion, see \S~\ref{section:crtran}). We also incorporate a moderate level of thermal conduction into our models. As our simulations show, thermal conduction delays the onset of the cooling catastrophe during the early stages of the ICM evolution, while the feedback heating by cosmic rays suppresses the cooling catastrophe quickly after it starts. 

The remainder of this paper is organized as follows. In \S~\ref{section:cosmicray} we describe cosmic-ray heating mechanisms and some physical assumptions we made in our models. In \S~\ref{section:equation} we write down the relevant differential equations of the intracluster medium.
In \S~\ref{section:evolution}, we present the results of a series of numerical hydrodynamic simulations that investigate the evolution of the ICM, starting from isothermal hydrostatic equilibrium. We conclude with a summary of our results in \S~\ref{section:conclusion}. The derivation of the time-dependent cosmic-ray equations is presented in the Appendix. In computing luminosity and angular diameter distances, we have rescaled observational results if the original paper used a different cosmology to the values $\Omega_{\text{m}}=0.3$, $\Omega_{\Lambda}=0.7$, $h=0.7$.

\section{Cosmic Ray Physics in Clusters of Galaxies}
\label{section:cosmicray}

\subsection{Cosmic-ray heating mechanisms}
\label{section:cosmicray1}

In this subsection, we quantify various cosmic ray energy loss mechanisms in clusters of galaxies, and their heating effects. For simplicity, we only consider relativistic protons in cosmic rays. Relativistic electrons are generally insufficient to heat the intracluster gas~\citep{1987MNRAS.225..851R}. 

\subsubsection{ Hydromagnetic waves}
\label{section:hydrowave}

The significance of the resonant interaction of cosmic rays with hydromagnetic waves has been recognized and discussed by many authors in various astrophysical contexts (e.g., \citet{1968Ap&SS...2..171M}; \citet{1971ApJ...170..265S}; \citet{1978ApJ...221L..29B}; \citet{1991ApJ...377..392L}; \citet{2004MNRAS.350.1174B}). During wave-particle resonance, the waves may be damped or grow exponentially, depending on the cosmic ray distribution function (see \citet{1968Ap&SS...2..171M}; \citet{2005ppfa.book.....K}, Chap. 12). If the distribution function of cosmic rays is sufficiently anisotropic due to their streaming motion (driven by the cosmic-ray pressure gradient; see \citet{2005ppfa.book.....K}, Chap. 12.6) with a streaming speed in excess of the local Alfv$\acute{\text{e}}$n speed, they will render ``forward" Alfv$\acute{\text{e}}$n waves propagating along the magnetic lines in the direction of the streaming unstable, while backward Alfv$\acute{\text{e}}$n waves are damped (\citet{2005ppfa.book.....K}, Chap. 12; see also \citet{1967ApJ...147..689L}, \citet{1969ApJ...156..445K}). During this quasi-linear cyclotron resonance (here also called the cosmic ray streaming instability), forward (and nearly forward) Alfv$\acute{\text{e}}$n waves grow fastest \citep[e.g.][]{1969ApJ...156..445K}, scatter the cosmic rays and reduce the cosmic-ray streaming speed to around the  Alfv$\acute{\text{e}}$n speed \citep[e.g.][]{1971ApJ...170..265S}. Cosmic rays are thus advected with these waves as scattering centers. 

These unstable Alfv$\acute{\text{e}}$n waves grow exponentially until saturated by nonlinear processes,
e.g., nonlinear Landau damping (see \citet{1973Ap&SS..24...31L}; \citet{1981IAUS...94..251C}; \citet{1984A&A...130...19V}; \citet{2005ppfa.book.....K}, Chap. 11.5) and are thus dissipated efficiently in the ionized thermal gas. Note that the decay of a forward Alfv$\acute{\text{e}}$n wave to a backward Alfv$\acute{\text{e}}$n wave and a forward sound wave is forbidden in the ICM where the sound speed $c_{\text{s}}$ is usually larger than the Alfv$\acute{\text{e}}$n speed $v_{\text{A}}$ (the requirement for the wave decay is $v_{\text{A}}>c_{\text{s}}$; see \citet{2005ppfa.book.....K}, Chap. 11.4 and \citet{1975MNRAS.173..245S}). Therefore, momentum and energy are transferred from relativistic protons to the waves and hence to the intracluster medium at the rates (per volume) $|\boldsymbol{\nabla} P_{\text{c}}|$ and $-(\boldsymbol{u}+\boldsymbol{v}_{\text{A}}) \boldsymbol{\cdot \nabla} P_{\text{c}}$, respectively (see Appendix A for the details; also see \citet{1971ApJ...163..503W}), where $P_{\text{c}}$ is the cosmic-ray pressure, $\boldsymbol{u}$ is the bulk velocity of the thermal plasma and $\boldsymbol{v}_{\text{A}}$ is the propagation velocity of the hydromagnetic waves relative to the plasma, which is equal to the local Alfv$\acute{\text{e}}$n velocity of the gas. The energy gained from the cosmic rays can accelerate and heat the gas \citep{1971ApJ...163..503W}. The increase in gas kinetic energy due to the work done by the CR pressure gradient is obviously $-\boldsymbol{u}\boldsymbol{\cdot \nabla} P_{\text{c}}$. Thus the cosmic-ray heating rate of the ICM due to the dissipation of hydromagnetic waves (hereinafter designated as ``cosmic-ray wave heating") is \citep{1971ApJ...163..503W, 1988ApJ...330..609B, 1991ApJ...377..392L}
\begin{equation}
\Gamma_{\text{wave}}=-\boldsymbol{v}_{\text{A}} \boldsymbol{\cdot \nabla} P_{\text{c}}\text{.} 
\end{equation}

We note that only Alfv$\acute{\text{e}}$n waves self-excited by the cosmic-ray streaming instability are considered in this paper. In reality, the MHD waves and turbulence in galaxy clusters may be much more complex. The MHD turbulence driven by cluster mergers may reaccelerate cosmic rays, which  has been studied by many authors to explain radio halos and hard X-ray tails in some galaxy clusters \citep[e.g.][]{2001ApJ...557..560P, 2004MNRAS.350.1174B, 2007MNRAS.378..245B}. These models usually assume an isotropic, homogeneous cosmic-ray phase space distribution function $f_{p}(p, t)$, and the cosmic rays are generally reaccelerated through the wave-particle resonance as long as $\partial f_{p}/\partial p < 0$ (the waves grow exponentially if $\partial f_{p}/\partial p > 0$; see, e.g., \citet{1968Ap&SS...2..171M}). In our model, the cosmic rays are mainly injected into the ICM by the central AGN and the cosmic ray pressure gradient is strong. In this case, the cosmic-ray streaming along the CR pressure gradient is significant, and the forward Alfv$\acute{\text{e}}$n waves self-excited by the cosmic-ray streaming instability should be the main MHD waves responsible for the cosmic-ray scattering \citep[see][]{2000PhRvL..85.4656C}. Note that cosmic ray streaming also depends on the details of the CR scattering off the small-scale MHD turbulence in the ICM. The details of the latter is still poorly understood at this point, and further observational and theoretical work is needed.  

It is worth noting, not only the ``drifted" cosmic-ray distributions, but also prolate ($p_{z} > p_{\perp}$) or oblate ($p_{z} < p_{\perp}$) CR distributions may generate Alfv$\acute{\text{e}}$n waves through cyclotron resonance (see \citet{2005ppfa.book.....K}, Chap. 12.3). \citet{2006MNRAS.373.1195L} recently proposed a new model for the generation of small-scale Alfv$\acute{\text{e}}$n waves through cyclotron resonance, where the prolate or oblate distributions of cosmic rays are driven by turbulent compressions of magnetic field.

\subsubsection{Coulomb interactions} 

The relativistic protons can transfer energy to the gas via Coulomb collisions with the ambient ionized gas. The heating rate due to Coulomb interactions of a fully ionized gas with a cosmic-ray particle with charge $Ze$, velocity $v=\beta c$ and kinetic energy $E$ is give by \citep{1994A&A...286..983M}
\begin{equation}
-\left(\frac{dE}{dt}\right)_{\text{C}}=4.96\times 10^{-19} Z^{2}\left(\frac{n_{\text{e}}}{\text{cm}^{-3}}\right)\frac{\beta ^{2}}{\beta ^{3}+x_{\text{m}}^{3}}\text{ergs s}^{-1}\text{ ,} 
\end{equation}
where $x_{\text{m}}=0.0286(T/2\times 10^{6} \text{K})^{1/2}$, $T$ and ${n_{\text{e}}}$ are the ambient electron temperature and number density. For simplicity, we neglect $x_{\text{m}}$ in our calculations, which is a good approximation for $\beta>0.15$ (i.e. $E \gtrsim10$ MeV) for a typical cluster temperature $T\sim 5$ keV.

\subsubsection{Hadronic collisions} 

The cosmic-ray protons interact hadronically with the ambient thermal gas and produce mainly $\pi^{+}$, $\pi^{-}$, $\pi^{0}$, provided their kinetic energy exceeds the threshold $E_{\text{thr}} = 282$ MeV for the reaction. The neutral pions decay after a mean lifetime of $9 \times 10^{-17}$ s into $\gamma$-rays ($\pi^{0} \rightarrow 2\gamma$), which may be detected by $\gamma$-ray observations with imaging atmospheric Cherenkov telescopes or the GLAST satellite\footnote{Gamma-ray Large Array Space Telescope (GLAST), homepage http://glast.gsfc.nasa.gov/}. One may thus test our model with the future observation of these $\pi^{0}$-decay induced $\gamma$-rays from galaxy clusters \citep[e.g., see][]{2007astro.ph..1033H}.
The charged pions decay into $e^{\pm}$ and neutrinos ($\pi^{\pm} \rightarrow \mu^{\pm} + \nu_{\mu}/\overline{\nu}_{\mu} \rightarrow e^{\pm} + \nu_{\text{e}}/\overline{\nu}_{\text{e}} + \nu_{\mu} + \overline{\nu}_{\mu}$). Since the limiting value of the inelasticity of these hadronic collisions is roughly $1/2$ \citep{1994A&A...286..983M}, the energy loss rate of a cosmic-ray proton due to pion production is approximately  \citep{2006astro.ph..3484E}
 \begin{equation}
\left(\frac{dE}{dt}\right)_{\text{h}}\approx -0.5n_{\text{N}} \sigma_{\text{pp}} \beta c E \theta(E-E_{\text{thr}})\text{,} 
\end{equation}
where $n_{\text{N}}=n_{\text{e}}/(1-\frac{1}{2}Y)$ is the target nucleon density in the ICM, $Y$ is the helium fraction, and $\sigma_{\text{pp}}$ is the pp cross section for the incident proton. We adopt an approximate value for $\sigma_{\text{pp}}$ ($\sigma_{\text{pp}}\approx 37.2$ mbarn) by using equation (69) of \citet{2006astro.ph..3484E} and assuming that the spectral index of the $\pi^{0}$-decay induced $\gamma$-ray spectrum is $2.5$.

To estimate the total energy loss rate of CRs due to Coulomb interactions and hadronic collisions, we need to determine the cosmic-ray energy spectrum. Galactic CR observations and many CR acceleration models suggest that the CR spectrum is a power-law in momentum \citep[see][for a review]{2002cra..book.....S}. However, in low energy regimes, the energy losses of cosmic rays are dominated by Coulomb interactions, which substantially flatten the spectrum. Similar to \citet{2006astro.ph..3484E}, we derive an approximate steady-state CR spectrum to estimate the total Coulomb and hadronic loss rates. Assuming that the cosmic rays are injected continuously and subject to energy losses through Coulomb and hadronic collisions, the cosmic ray spectrum in a homogeneous environment follows the evolution equation:
\begin{equation}
\frac{\partial f_{E}(E, t)}{\partial t}+\frac{\partial}{\partial E}\left(f_{E}(E, t)\frac{dE}{dt}\right)=Q_{\text{E}}(E)
\text{ ,} 
\end{equation}
where the cosmic-ray spectrum $f_{E}(E, t)$ is defined as
\begin{equation}
f_{E}(E, t)=\frac{dN}{dE\text{ }dV} =4\pi p^{2}f_{p}(\boldsymbol{x}, \boldsymbol{p}, t)\frac{dp}{dE} \text{,} 
\end{equation}
and 
\begin{equation}
\frac{dE}{dt}=\left(\frac{dE}{dt}\right)_{\text{C}}+\left(\frac{dE}{dt}\right)_{\text{h}}
\end{equation}
is the energy loss rate due to Coulomb and hadronic collisions. For simplicity, here the cosmic-ray phase space distribution function $f_{p}(\boldsymbol{x}, \boldsymbol{p}, t)$ is assumed to be isotropic in momentum space and we approximate hadronic losses as continuous, rather than impulsive. Assuming that the CR injection spectrum is a power law in momentum, i.e. 
\begin{equation}
Q_{p}(p)\propto p^{-\alpha}\theta(p-p_{l})\text{,} \label{ispectrum}
\end{equation}
where $\theta(x)$ is the Heaviside step function and $p_{l}$ is the momentum cutoff, we can get the CR injection spectrum $Q_{\text{E}}(E)$:
\begin{equation}
 Q_{\text{E}}(E)\propto (E+E_{0})[E(E+2E_{0})]^{-(\alpha+1)/2}\theta(E-E_{l})     \text{ ,} 
\end{equation}
where $E_{l}$ is the energy cutoff and $E_{0}=938$ MeV is the proton rest energy. We find the asymptotic steady-state spectrum by assuming negligible hadronic and Coulomb losses in the low and high energy regimes, respectively:
\begin{equation}
f_{E}(E) =A_{\text{cr}}
\begin{cases}
(E/E_{\ast})^{-\alpha} \quad  &\text{for } E \gg E_{\ast} \text{,}\\
(E/E_{\ast})^{-(\alpha-2)/2} \quad &\text{for } E_{l}<E \ll E_{\ast}\text{,}
\end{cases}
\end{equation}
where $A_{\text{cr}}$ is the normalization factor which can be determined in terms of the energy density ($E_{\text{c}}$) of cosmic rays, and the cross-over energy $E_{\ast} \approx 706$ MeV depends on the ratio of the Coulomb to hadronic loss rates. We note that a similar asymptotic stationary CR spectrum has also been derived by \citet{2004MNRAS.350.1174B} and \citet{2006astro.ph..3484E}. We adopt a simple analytic approximation for the steady-state CR spectrum with the same asymptotic behaviors:
\begin{equation}
f_{E}(E) =\frac{A_{\text{cr}}\theta(E-E_{l})}{(E/E_{\ast})^{\alpha} +(E/E_{\ast})^{(\alpha-2)/2}} \text{ .} \label{spectrum}
 \end{equation}

Using the approximate steady-state spectrum (eq. [\ref{spectrum}]), we get the overall Coulomb loss rate of cosmic rays
\begin{align}
&\Gamma_{\text{C}} = \int_{E_{l}}f_{\text{E}}(E) \left(\frac{dE}{dt}\right)_{\text{C}}dE \notag \\
&= - 1.65 \times 10^{-16} \left(\frac{n_{\text{e}}}{\text{cm}^{-3}}\right) \left(\frac{E_{\text{c}}}{\text{ergs cm}^{-3}}\right)\text{ergs s}^{-1}\text{ cm}^{-3}  \text{ ,}  
\end{align}
where $E_{c}$ is the energy density in cosmic rays, and we adopt $\alpha = 2.5$ and $E_{l} = 10$ MeV. The loss rate $\Gamma_{\text{C}}$ depends slightly on the value of $E_{l}$, but not very sensitively, since in the low energy regimes $f_{\text{E}}(E)$ is quite flat (i.e., the CR spectral index, $(\alpha-2)/2$, is quite small). Similarly, the overall hadronic loss rate per volume is 
\begin{align}
&\Gamma_{\text{h}} = \int_{E_{\text{thr}}}f_{\text{E}}(E) \left(\frac{dE}{dt}\right)_{\text{h}}dE \notag \\
&=  -5.86 \times 10^{-16} \left(\frac{n_{\text{e}}}{\text{cm}^{-3}}\right) \left(\frac{E_{\text{c}}}{\text{ergs cm}^{-3}}\right)\text{ergs s}^{-1}\text{ cm}^{-3}  \text{ .}  
\end{align}
We note that CR Coulomb losses are usually sub-dominant with respect to hadronic losses if we use the stationary CR spectrum (eq. [\ref{spectrum}]) with $\alpha \lesssim 3.7$ to calculate the CR energy loss rates. For higher values of $\alpha$ or using the CR injection spectrum (eq. [\ref{ispectrum}] with $\alpha \gtrsim 2$) instead of the stationary spectrum, Coulomb losses usually dominate.  
 
Therefore, the total cosmic-ray energy loss rate due to Coulomb and hadronic collisions is $\Gamma_{\text{loss}}=\Gamma_{\text{h}}+\Gamma_{\text{C}}=-\zeta_{\text{c}}n_{\text{e}}E_{\text{c}}$ where $\zeta_{\text{c}} =7.51 \times 10^{-16}\text{ cm}^{3}\text{ s}^{-1}$ is the rate coefficient for collisional energy loss of the cosmic rays. While the CR energy lost in Coulomb interactions heats the ICM, most of the CR energy lost in hadronic collisions tends to escape via gamma rays and neutrinos. During hadronic collisions, a small fraction ($\sim 1/6$, see \citet{1994A&A...286..983M}) of the inelastic energy goes into secondary electrons. While the cooling of high-energy electrons ($\gamma \gtrsim 10^{3}$) is dominated by synchrotron and inverse Compton losses, most of the mildly relativistic ($\gamma \lesssim 200$) electrons will heat the ICM by Coulomb interactions with the plasma electrons and through plasma oscillations and excitation of Alfv$\acute{\text{e}}$n waves \citep[e.g.][]{1979ApJ...227..364R}. Since the spectrum of secondary electrons is dominated by mildly relativistic electrons, here we assume that these secondary electrons lose most of their energy through thermalization and thus heat the ICM. Therefore, the ICM is heated by cosmic rays through Coulomb and hadronic collisions (hereinafter designated as ``CR collisional heating'') at a rate of $\Gamma_{\text{coll}}=-\Gamma_{\text{C}}-\Gamma_{\text{h}}/6=\eta_{\text{c}}n_{\text{e}}E_{\text{c}}$, where $\eta_{\text{c}} =2.63 \times 10^{-16}\text{ cm}^{3}\text{ s}^{-1}$. 

Note that the real CR spectrum in galaxy clusters may differ substantially with our simple steady state spectrum, since our calculation neglects the influence of CR transport and energy losses due to the generation of hydromagnetic waves. The detailed shape of the spectrum has relatively little impact on hadronic losses. The Coulomb heating rate does depend on the CR spectrum, but even if these effects are included, the CR spectrum is still strongly flattened at low energies due to strong Coulomb losses. The heating rate from the steady-state CR spectrum (which is quickly reached since $t_{\rm Coulomb} \ll t_{\rm cluster}$ at low energies) is an order of magnitude less than the heating rate calculated from the injection spectrum (e.g., equation (3) in \citet{2006MNRAS.366L..35N}), which is only a short transient. As shown in \S~\ref{section:evolution2}, CR collisional heating ($\eta_{\text{c}} =2.63 \times 10^{-16}\text{ cm}^{3}\text{ s}^{-1}$), is negligible comparing to CR wave heating. Thus, our results should not change appreciably if a new collisional heating rate calculated with a more realistic CR spectrum is used instead. 

\subsection{Cosmic-ray propagation and assumptions}
\label{section:crtran}

In our models, cosmic rays are assumed to be primarily injected into the ICM by AGN-produced radio cocoons or bubbles. Cosmic rays then propagate along the magnetic field lines, which, for simplicity, are assumed to be mainly radial on a large scale. Such a magnetic field structure could be created by the bubbles themselves: magnetic fields stretch out behind the bubble and become approximately parallel to the (radial) direction of motion \citep{2007astro.ph..3801R}. Cosmic rays can thus be viewed as streaming radially on a large scale. If so, cosmic ray propagation can be described approximately by a simplified one-dimensional model with spherical symmetry. In the future, it will be interesting to consider more realistic 3D hydromagnetic simulations, which (among other effects) could potentially take into account the effects of cross-field diffusion of CRs.

The time-dependent cosmic-ray equations governing CR transport and CR energy loss due to the generation of Alfv$\acute{\text{e}}$n waves are derived in Appendix A. As shown in the previous subsection, cosmic rays in the ICM can also lose their energy through Coulomb and hadronic collisions with the ambient thermal gas, at a rate $\Gamma_{\text{loss}}=\zeta_{\text{c}}n_{\text{e}}E_{\text{c}}$. Taking CR collisional losses into account, the net CR source function $\bar{Q}$ in equation (\ref{firstenergy2}) may be written as $\bar{Q}=Q_{\text{c}}-\zeta_{\text{c}}n_{\text{e}}E_{\text{c}}$, where $Q_{\text{c}}$ is the source (injection) function of cosmic-ray energy. 

Assuming spherical symmetry, the CR energy equation (\ref{firstenergy2}) may be rewritten as 
\begin{equation}
\frac{\partial E_{\text{c}}}{\partial t} =(\gamma_{\text{c}}-1) 
(u+v_{\text{A}})\frac{\partial E_{\text{c}}}{\partial r}-
\zeta_{\text{c}}n_{\text{e}}E_{\text{c}}
- \frac{1}{r^{2}}\frac{\partial (r^{2}F_{\text{c}})}{\partial r} + Q_{\text{c}} \text{,} \label{cosmic1}
\end{equation}
where $u$ is the radial velocity of the thermal plasma, $ v_{\text{A}}$ is the local Alfv$\acute{\text{e}}$n speed of the thermal gas, $F_{\text{c}}$ is the cosmic-ray energy flux, and $\gamma_{\text{c}}=4/3$ is the adiabatic index for the cosmic rays. The cosmic-ray energy flux $F_{\text{c}}$ is (see eq. [\ref{energyflux2}]) 
\begin{equation}
F_{\text{c}}=\gamma_{\text{c}}(u+v_{\text{A}})E_{\text{c}}-\kappa_{\text{c}}\frac{\partial E_{\text{c}}}{\partial r} \text{,} \label{cosmic2} 
\end{equation}
where $\kappa_{\text{c}}$ is the diffusion coefficient of cosmic rays. The two terms in equation (\ref{cosmic2}) clearly represent the spread of cosmic rays by advection and diffusion respectively.
  
Magnetic fields in the ICM have been measured using a variety of techniques, such as Faraday rotation measurements and studies of  synchrotron radiation and inverse Compton X-ray emission \citep[see][for a review]{2002ARA&A..40..319C}. These measurements imply that the ICM of most clusters is substantially magnetized, with a typical field strength of order 1 $\mu$G with high areal filling factors out to Mpc radii. In the cores of cool core clusters, these measurements also suggest that magnetic field strength is typically much higher, up to 10s of $\mu$G \citep[e.g.][]{2001MNRAS.324..842A}. 
 
The magnetic field is usually not uniform and its distribution in galaxy clusters is still far from clear. For definiteness, similar to \citet{jubelgas06}, we assume that the magnetic pressure ($P_{\text{B}}=B^{2}/(8\pi)$) is a fixed fraction of the thermal pressure ($P_{\text{g}}$), which corresponds to
\begin{equation}
B\propto n_{\text{e}} ^{1/2}T^{1/2}
\text{.}
\label{bfield}
\end{equation}
In the rest of this paper, we assume that $P_{\text{B}}/P_{\text{g}}=0.06$, which corresponds to $B \sim 10-20$ $\mu$G in the central regions of the cluster in the final steady state. Such magnetic field amplitudes have been found in the cores of cool core clusters through the analysis of Faraday Rotation measure (RM) maps \citep[e.g.][]{2003A&A...412..373V}. We neglect pressure support from magnetic fields, since this small level of magnetic field pressure will not substantially change our main results. We also neglect the pressure of the hydromagnetic waves, which should at most be comparable to the pressure of the main field.

The diffusion coefficient of cosmic rays in the intracluster medium, which depends on the frequecy of pitch-angle scattering and the cosmic-ray momentum spectrum, is highly uncertain and may vary substantially between different parts of galaxy clusters. For definiteness, here we follow a simple treatment given by \citet{jubelgas06} and discuss the dependence of our model on radial profiles of $\kappa_{\text{c}}$ in \S~\ref{section:evolution2}. In Kolmogorov-like MHD turbulence, the parallel diffusivity is expected to scale as
\citep{2003A&A...399..409E}
\begin{equation}
\kappa_{\text{c}} \propto l_{\text{B}}^{2/3}B^{-1/3}
\text{,}
\end{equation}
where $l_{\text{B}}$ gives a characteristic length scale for the magnetic field of strength $B$. Here we have ignored the weak dependence of $\kappa_{\text{c}}$ on the cosmic-ray energy distribution. $l_{\text{B}}$ is fairly unknown in galaxy clusters. For definiteness, we assume that  $l_{\text{B}}$ scales with the local Jeans length \citep{jubelgas06}, and thus obtain: 
\begin{equation}
\kappa_{\text{c}} =  \kappa_{0} \left(\frac{ n_{\text{e}}}{0.12 \text{ cm}^{-3}}\right)^{-1/2}\left(\frac{T}{1.6\text{  keV}}\right)^{1/6} \text{ cm}^{2} \text{ s}^{-1}\text{,}
\label{diff_coef}
\end{equation}
where we fixed the normalization $\kappa_{0}$ by assuming that $\kappa_{\text{c}}$ at the center of the fiducial cluster Abell 2199 equals to the estimated diffusion coefficient along the magnetic field lines in the interstellar medium of our own Galaxy: $\kappa_{0}=\kappa_{c,\text{ ISM}} \approx 3 \times 10^{28}$ cm$^{2}$ s$^{-1}$ \citep{1990acr..book.....B}. The values of $n_{\text{e}}(r_{\text{min}})  =0.12\text{ cm$^{-3}$}$ and $T(r_{\text{min}})  =1.6\text{ keV}$ at the inner boundary $r_{\text{min}}=1$ kpc are obtained directly from Chandra observations of the cluster Abell 2199 \citep{2002MNRAS.336..299J}. Obviously, this estimate is highly uncertain, but it is consistent with the approximate bound of $\kappa_{\text{c}} \lsim 7.5 \times 10^{29}r_{\text{lobe},5}^{2}t_{\text{lobe},7}^{-1}$ cm$^{2}$ s$^{-1}$ required for large cavities of radii $r_{\text{lobe}} \sim 5r_{\text{lobe},5}$ kpc to remain stable to the diffusion of cosmic rays in time $t_{\text{lobe}} \sim 10^{7}t_{\text{lobe},7}$ yr, as required by observations \citep{2007ApJ...660.1137M}, as well as theoretical estimates \citep{2003A&A...399..409E}. 

\subsection{Injection of cosmic rays by a central AGN}
\label{section:AGN_injection}

The injection of CRs is specified by the cosmic-ray source function $Q_{\text{c}}$, which depends on the spatial profile of CR injection from AGN activity, supernovae, structure formation shocks and merger shocks. The form of $Q_{\text{c}}$ is far from clear. In this paper, we only consider CR injection from the central AGN in clusters, via jet-ICM interactions, which create buoyant bubbles filled with cosmic rays. When the bubbles rise, they inject cosmic rays into the ICM through diffusion or the shredding of the bubbles by Rayleigh-Taylor (RT) and Kelvin-Helmholtz (KH) instabilities. This is an efficient means of transporting cosmic rays from the AGN out to large distances: the buoyancy timescale is typically comparable to (at most several times) the sound crossing time $t_{\rm sc} \sim 10^{8} r_{100} c_{s,1000}^{-1}$yr for a radius $r \sim 100 r_{100}$ kpc and sound speed $c_{\rm s} \sim 1000 c_{s,1000} \, {\rm km \, s^{-1}}$ (e.g., see table 3 in \citet{2004ApJ...607..800B}), whereas the CR diffusion time along field lines is $t_{\rm diffusion} \sim 3 \times 10^{11} r_{100}^{2} \kappa_{\text{c},28}^{-1}$yr (where the diffusivity $\kappa_{\text{c}} \sim 10^{28} \kappa_{\text{c},28}$ cm$^{2}$ s$^{-1}$), too long to affect the thermal state of gas at the cooling radius. Efficient advection of cosmic rays decreases this timescale by about an order of magnitude (see \S \ref{section:evolution2}), still too slow. Bubble transport is a key ingredient of our model: if excluded, cosmic-ray transport is too slow to allow significant heating (e.g., models of \citet{1988ApJ...330..609B}). 

AGN activity is likely to be intermittent on a timescale of order the Salpeter time $t_{\rm S} \sim 10^{7}$ yr, and possibly as short as $t_{\text{i}} \sim 10^{4}-10^{5}$ yr (\citet{2002ApJ...581..223R}, hereafter RB02; \citet{1997ApJ...487L.135R}), which is much shorter than the bubble rising time. Note that the bubble rise time is usually much shorter than the gas cooling time (RB02). It is thus justifiable to assume that the CR injection into the ICM from the buoyant bubbles, which are produced by a succession of AGN outbursts, can be treated in a time-averaged sense. Since CR transport timescales are much shorter than thermal timescales, we assume that the cosmic rays are injected into the ICM instantaneously and neglect any delay between central AGN activity and the cosmic-ray injection (similar to instantaneous AGN mechanical heating models, e.g., RB02, \citet{2003ApJ...587..580B}).
 
The rate at which bubbles are disrupted is highly uncertain, since the nature of the physical mechanism which protects them from RT and KH instabilities---perhaps an ordered magnetic field at the bubble surface, or thermal conduction/plasma viscosity \citep{2005MNRAS.359..493K,2005MNRAS.357..242R}, or even the initial deceleration and drag on the bubble-ICM interface during inflation \citep{2006MNRAS.371.1835P}--is not well understood. Likewise, the diffusion rate of CRs out of the bubbles is highly uncertain, particularly given the unknown magnetic field topology at the bubble interface \citep[see][for recent simulations]{2007arXiv0705.3235R}. Nonetheless, the observational fact remains that bubbles are seen to survive intact to large radii (with an average projected radius of $\sim 20$ kpc in a sample of 16 clusters \citep{2004ApJ...607..800B}). We hence parametrize our ignorance of the bubble disruption rate by simply assuming that the spherically integrated CR energy flux in the buoyant bubbles is a power law with radius:
\begin{equation}
L_{\text{bubble}} \sim  -\epsilon \dot{M}_{\text{in}}c^{2} \left(\frac{r}{r_{0}}\right)^{-\nu}\quad \text{for } r>r_{0} \text{,} \label{lbubble}
\end{equation}
where $\epsilon$ is the efficiency with which the rest-mass energy of the ICM cooling flow is converted into the cosmic-ray energy in the bubbles, $\dot{M}_{\text{in}}$ is the mass accretion rate across the inner boundary of the simulation, $c$ is the speed of light, $\nu$ is a positive constant, and $r_{0}$ is a characteristic radius where the bubbles are created. The decline of $L_{\rm bubble}$ with radius reflects the transfer of CRs to the ICM. In particular, the cosmic-ray energy injection rate into the ICM per unit volume is given by:
\begin{align}
Q_{\text{c}} = -\nabla \cdot & {\bf F}_{\rm bubble} \sim -\frac{1}{4\pi r^{2}}\frac{\partial L_{\text{bubble}}}{\partial r}\left[1-e^{-(r/r_{0})^{2}}\right]\notag \\
&\sim -\frac{\nu \epsilon\dot{M}_{\text{in}}c^{2}}{4\pi r_{0}^{3}}\left(\frac{r}{r_{0}}\right)^{-3-\nu}
\left[1-e^{-(r/r_{0})^{2}}\right]
\text{,} 
\label{cosmicsource}
\end{align}
where $F_{\rm bubble}= L_{\rm bubble}/(4 \pi r^{2})$, and we have introduced an inner injection cutoff term, which reflects the finite radius $r \sim r_{0}$ at which bubbles are injected; it is very similar to the AGN heating cutoff term in RB02. Here $r_{0}$ was taken to be $20$ kpc. Inclusion of this term allows us to apply equation (\ref{cosmicsource}) at all radii in our simulations. 

Besides CR injection into the ICM, bubbles also lose energy by expansion as they rise. Therefore, strictly speaking, $L_{\text{bubble}}$ defined in equation (\ref{lbubble}) is only a means of parametrizing the spatial distribution of the CR injection rate (eq. [\ref{cosmicsource}]), and is generally different from the total energy flux of cosmic rays. We can include the PdV work from bubble expansion in equation (\ref{lbubble}):
\begin{equation}
L_{\text{bubble}} \sim  - \dot{M}_{\text{in}} c^{2}  \left[ \epsilon \left(\frac{r}{r_{0}}\right)^{-\nu}+ \mathcal{F}_{\text{pdv}}\right] 
 \quad \text{for } r>r_{0} \text{.} \label{lbubble2}
\end{equation}
As before, the first term on the right-hand side of equation (\ref{lbubble2}) represents the cosmic-ray injection into the ICM, while the second term $\mathcal{F}_{\text{pdv}}$ represents the PdV work done on the ICM during the bubble expansion. The AGN feedback efficiency $\epsilon_{\text{tot}}$ is:
\begin{equation}
\epsilon_{\text{tot}}\sim-\frac{L_{\text{bubble}}(r_{0})}  {\dot{M}_{\text{in}} c^{2}}.   
\end{equation}
If the bubble is nearly in pressure equilibrium with the ICM, $\mathcal{F}_{\text{pdv}}$ satisfies the equation:
\begin{equation}
(\gamma_{\text{c}}-1)P_{\text{g}}\frac{d}{dr}\left[\frac{\epsilon (r/r_{0})^{-\nu}+ \mathcal{F}_{\text{pdv}}}{P_{\text{g}}}\right] 
=-\frac{d\mathcal{F}_{\text{pdv}}}{dr}\text{.}\label{lbubble3}
\end{equation}
If we further assume that $P_{\text{g}}$ scales approximately with $r^{-\tau}$ ($\tau>0$), the solution of equation (\ref{lbubble3}) is:  
\begin{equation}
\mathcal{F}_{\text{pdv}}\sim (\epsilon_{\text{tot}}-\epsilon)\left(\frac{r}{r_{0}}\right)^{-\tau/\omega}+\mathcal{G}(r)\text{,}
\end{equation}
where $\omega=\gamma_{\text{c}}/(\gamma_{\text{c}}-1)$, and
\begin{equation}
\mathcal{G}(r)=
\begin{cases}
\frac{\epsilon(\tau-\nu)}{\omega \nu-\tau}\left[\left(\frac{r}{r_{0}}\right)^{-\nu}-\left(\frac{r}{r_{0}}\right)^{-\tau/\omega}\right]\quad &\text{if }\nu\neq\frac{\tau}{\omega}\\
-\frac{\epsilon\tau(\omega-1)}{\omega^{2}}\left(\frac{r}{r_{0}}\right)^{-\tau/\omega}\text{ln}\left(\frac{r}{r_{0}}\right)&\text{if }\nu=\frac{\tau}{\omega}\text{.}
\end{cases} 
\end{equation}

\begin{figure}
\includegraphics[width=0.45\textwidth]{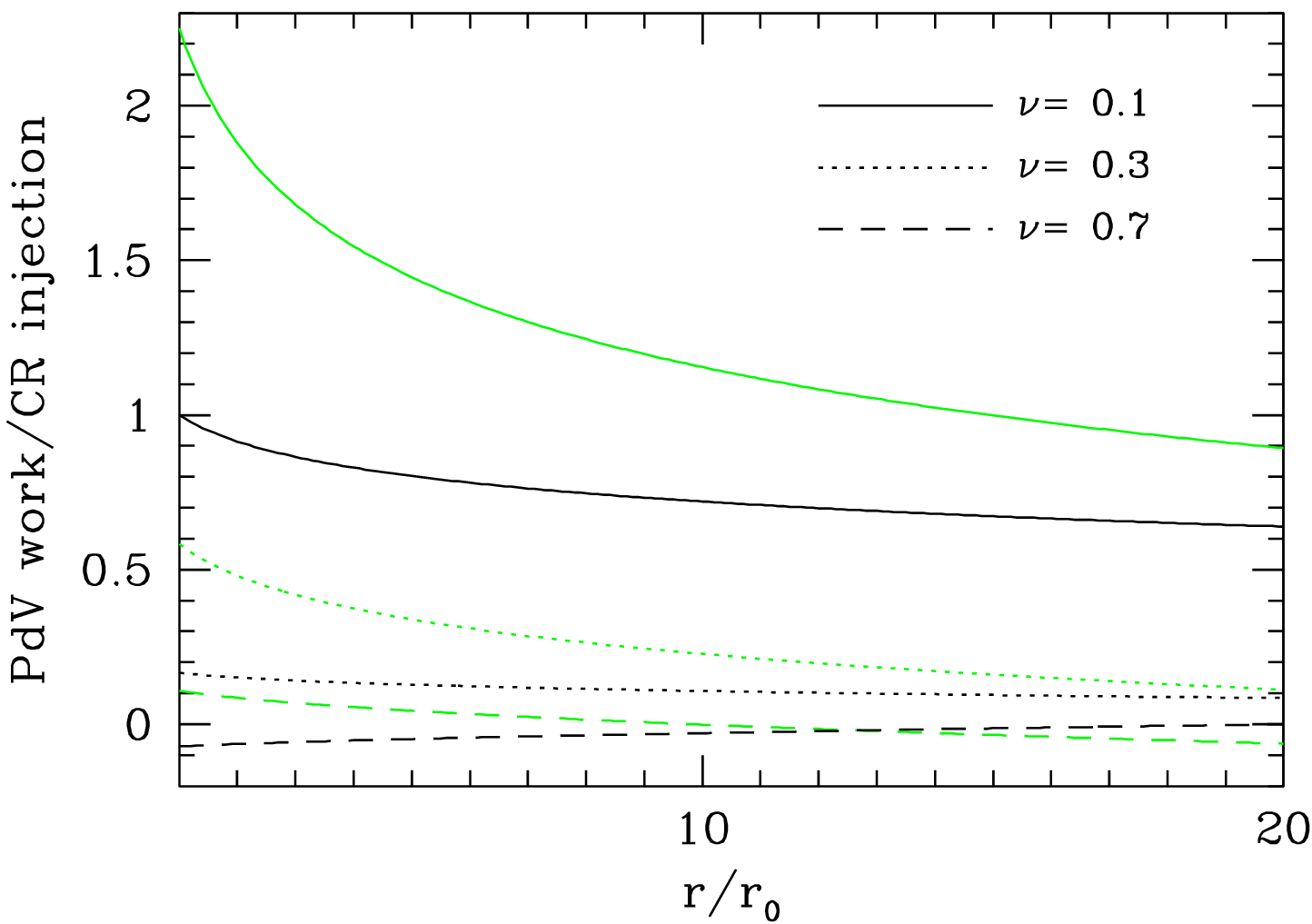}
\caption{The ratio of the bubble PdV work rate to the CR injection rate, plotted as a function of radius ($\gamma_{\text{c}}$ is taken to be $4/3$). For each line type, the upper line (green) corresponds to $\tau=1$, while the lower one (black) corresponds to $\tau=0.5$. Note that, in the central regions of the cluster, the bubble expansion is less important for smaller values of $\tau$. A positive value of the PdV work rate means that the bubble expands, while a negative value corresponds to the bubble contraction.}
 \label{pdvcr}
 \end{figure}

The AGN mechanical heating model of RB02 assumes no cosmic ray leakage from the bubbles and finds that $\mathcal{F}_{\text{pdv}}$ scales with $P_{\text{g}}^{(\gamma_{\text{c}}-1)/\gamma_{\text{c}}}$, which is the solution of equation (\ref{lbubble3}) when $\epsilon=0$. On the other hand, in this paper, we consider the heating of the ICM by the cosmic rays leaked from these AGN-produced bubbles, and assume that $\epsilon \sim \epsilon_{\text{tot}}$. In this case, the effect of the bubble expansion or contraction is described by the variation of $\mathcal{G}(r)$ with radius. For $0<\nu<\tau$, $\mathcal{G}(r)$ first decreases and then increases with radius, which means that the bubble expands first and contracts later. For $\nu>\tau$, the bubble contracts first and expands later. We are not interested in the models with high values of $\nu$ ($\nu\gtrsim 1.0$), where the cosmic rays are essentially dispersed into the ICM at the cluster center. When $\nu$ is very small ($\nu\lsim 0.1$), the CR injection is less important and the bubble PdV work (the model of RB02) dominates. In this paper, we are interested in the models with a moderate level of the bubble disruption ($0.1 \lsim \nu \lsim 1.0$), where the CR injection into the ICM dominates and hence the CR heating of the ICM may be significant. Figure \ref{pdvcr} shows the ratio of the bubble PdV work rate to the CR injection rate for different values of $\nu$ and $\tau$ ($\tau\sim 0.5-1.0$ in the central regions of the cluster Abell 2199). As can be clearly seen, bubble expansion is subdominant for $\nu=0.3$ or $0.7$, and becomes comparable to the cosmic ray injection only for very flat disruption profiles, $\nu=0.1$, reverting to the case studied by RB02. Hence, we shall neglect bubble expansion in this paper.   

The cosmic ray injection described by equations (\ref{lbubble}) and (\ref{cosmicsource}) is obviously simplified, and would be worth refining in more detail once more is known both observationally and theoretically about bubble disruption. Nonetheless, our model should be relatively robust to the details of the bubble disruption profile. For instance, as we show in \S~\ref{section:evolution2}, our model is very robust to the value of the model parameters (i.e., no fine-tuning of the model parameters $\epsilon$, $\nu$ needed).  
 
\section{Basic Equations}
\label{section:equation}

In our spherically symmetric model of galaxy clusters, the dark matter distribution is given by 
a Navarro-Frenk-White (NFW) profile \citep*{1997ApJ...490..493N}:
\begin{equation}
\rho_{\text{DM}}(r)=\frac{M_{0}/2\pi}{r(r+r_{\text{s}})^{2}}\text{,}  
\end{equation}
where $r_{\text{s}}$ is  the standard scale radius  of the NFW profile and $M_{0}$ is a characteristic mass.

In this paper, we will take the cooling flow cluster Abell 2199 as our fiducial model to study cosmic-ray heating in galaxy clusters. For this cluster, the parameters of the NFW profile are
$M_{0} = 3.8 \times 10^{14}$ M$_{\sun}$, $r_{\text{s}}=390$ kpc \citep*{2003ApJ...582..162Z}.
Since we are interested in the cluster within the cooling radius, the central cD galaxy NGC 6166 
may be dynamically important as well. We adopt a King profile with core radius and 
one-dimensional velocity dispersion of $r_{\text{g}}=2.83$ kpc and $\sigma=200 \text{ km s}^{-1}$, respectively, for the density distribution of NGC 6166 \citep{2002ApJ...576..720K}:
\begin{equation}
\rho_{cD}(r)=\frac{\rho_{0}}{[1+(r/r_{\text{g}})^{2}]^{3/2}}\text{,}   
\end{equation}
where $\rho_{0}$ is the central density,
\begin{equation}
 \rho_{0}=\frac{9\sigma ^{2}}{4\pi G r_{\text{g}}^{2}}\text{ .}   
\end{equation}

The intracluster medium is subject to radiative cooling, thermal conduction and cosmic-ray heating. The governing hydrodynamic equations are
\begin{equation}
\frac{\partial \rho}{\partial t} + \boldsymbol{\nabla \cdot}(\rho \boldsymbol{u}) = 0 \text{ ,}\label{hydro11}
\end{equation}
\begin{equation}
\frac{\partial \boldsymbol{S}}{\partial t} + \boldsymbol{\nabla \cdot}(\boldsymbol{Su})= -\boldsymbol{\nabla} P_{\text{g}}-\boldsymbol{\nabla} P_{\text{c}}-\rho \boldsymbol{\nabla} \Phi \text{ ,}\label{hydro21}
\end{equation}
\begin{align}
\frac{\partial E_{\text{g}}}{\partial t} +\boldsymbol{\nabla \cdot}(E_{\text{g}}\boldsymbol{u})=&-P_{\text{g}}\boldsymbol{\nabla \cdot u} - \boldsymbol{\nabla \cdot F}-n_{\text{e}}^{2}\Lambda(T) 
+\eta_{\text{c}}n_{\text{e}}E_{\text{c}}\notag  \\
&-
\boldsymbol{v}_{\text{A}} \boldsymbol{\cdot \nabla}P_{\text{c}}\text{ ,}\label{hydro31}
\end{align}
where $\rho $ is the gas density, $P_{\text{g}}$ is the gas pressure, $E_{\text{g}}$ is the gas energy density, $ \boldsymbol{S}= \rho \boldsymbol{u}$ is the gas momentum vector, $P_{\text{c}}=(\gamma_{\text{c}}-1)E_{\text{c}}$ is the cosmic-ray pressure and $\Phi$ is the gravitational potential, which is obtained by solving Poisson's equation:
\begin{equation}
\nabla ^{2} \Phi= 4\pi G (\rho_{\text{DM}}+ \rho_{\text{cD}})\text{,} \label{gravpotent}
\end{equation}
where we have neglected the self-gravity of the intracluster medium (ICM). Hence, $\Phi$ can be written as $\Phi=\Phi_{\text{DM}}+\Phi_{\text{cD}}$, where 
\begin{equation}
\Phi_{\text{DM}}= -\frac{2GM_{0}}{r_{\text{s}}} \frac{\text{ln}(1+r/r_{\text{s}})}{r/r_{\text{s}}}    
\end{equation}
is the cluster dark matter potential and
\begin{equation}
\Phi_{\text{cD}}= -4\pi G\rho_{0}r_{\text{g}}^{2} \frac{\text{ln}(r/r_{\text{g}}+\sqrt{1+(r/r_{\text{g}})^{2}})}{r/r_{\text{g}}}    
\end{equation}
is the gravitational potential contributed by the central cD galaxy. 
In our spherically symmetric model, we define the gravitation acceleration $g=d\Phi/dr$.

We adopt the ideal gas law,
\begin{equation}
 P_{\text{g}}=\frac{\rho k_{\text{B}} T}{\mu m_{\mu}}=\frac{\mu _{\text{e}}}{\mu}n_{\text{e}}k_{\text{B}}T\text{,}
\end{equation}
where $k_{\text{B}}$ is Boltzmann's constant, $m_{\mu}$ is the atomic mass unit, and $\mu$ and $\mu_{\text{e}}$ are the mean molecular weight per thermal particle and per electron, respectively. We assume that the gas is fully ionized with hydrogen fraction $X=0.7$ and helium fraction $Y=0.28$ \citep{2003ApJ...582..162Z}, so that  $\mu=0.62$ and $\mu_{\text{e}}=1.18$. We use an analytic fit \citep*{2001ApJ...546...63T} to the cooling function based on calculations by \citet{1993ApJS...88..253S},
\begin{align}
  n_{\text{e}}^{2}\Lambda(T)  = & 1.0 \times 10^{-22}\left[C_{1}\left(\frac{k_{\text{B}}T}{\text{keV}}\right)^{\delta_{1}}+C_{2}\left(\frac{k_{\text{B}}T}{\text{keV}}\right)^{\delta_{2}}+C_{3}\right] \notag  \\
  &\times \left(\frac{n_{\text{i}}}{\text{cm}^{-3}}\right) \left(\frac{n_{\text{e}}}{\text{cm}^{-3}}\right)  \text{ ergs cm$^{-3}$ s$^{-1}$}\text{ ,} \label{hydro2}
\end{align}
where $n_{\text{i}}$ is the ion number density. For an average metallicity $Z=0.3Z_{\sun}$, the constants are $\delta_{1}=-1.7$, $\delta_{2}=0.5$, $C_{1}=8.6\times 10^{-3}$, $C_{2}=5.8\times 10^{-2}$ and $C_{3}=6.3\times 10^{-2}$, and we can approximate $n_{\text{i}}n_{\text{e}}=0.704(\rho/m_{\text{p}})^{2}$, where $m_{\text{p}}$ is the proton mass. We manually truncate the cooling below a minimum temperature of $0.03$ keV, since equation (\ref{hydro2}) is only valid for $k_{\text{B}} T>0.03$ keV  \citep*{2001ApJ...546...63T}.

In equation~(\ref{hydro31}), $\boldsymbol{F}$ is the heat flux due to electron conduction,
 \begin{equation}
  \boldsymbol{F}  = 
  -f\kappa_{\text{Sp}} 
  \boldsymbol{\nabla}T \text{,} \label{hydro3}
\end{equation}
where $f$ ($0\leq f \leq 1$) is a conductivity reduction factor due to magnetic field suppression and $\kappa_{Sp}$ is the classical Spitzer conductivity \citep*{1962pfig.book.....S},
 \begin{equation}
  \kappa_{\text{Sp}}= \frac{1.84 \times 10^{-5}}{\text{ln} \lambda }T^{5/2} \text{ ergs s$^{-1}$ }\mathrm{K}^{-7/2}\text{ cm} ^{-1}\text{,}
\end{equation}
where $\text{ln} \lambda \sim 37$ is the usual Coulomb logarithm. For simplicity, here we assume that $f$ is constant throughout the cluster. In real clusters, heat transport may be much more complex, depending on the plasma magnetization and turbulence driving. Heat transport through turbulent fluid motions may also need to be taken into account \citep[see][]{2006ApJ...645L..25L}. Since at present there is no consensus on the nature of conductivity in a turbulent magnetized plasma, we adopt the same assumption of Spitzer conductivity (with a constant suppression factor) that most authors do. We do show that (unlike others) no fine-tuning of $f$ is necessary in our models. 
 
\begin{table}
 \centering
 \begin{minipage}{80mm}
  \renewcommand{\thefootnote}{\thempfootnote} 
  \caption{List of Simulations.}
  \label{listtable}
  \begin{tabular}{@{}llcccc}
  \hline
          Run & Heating  &  {$ f$\footnote{Conductivity suppression factor relative to the Spitzer value.}} & {$\epsilon$\footnote{Efficiency of cosmic ray injection due to accretion-triggered AGN activity. See eqs. (\ref{lbubble}) and (\ref{cosmicsource}).} }&{$\nu$\footnote{See eqs. (\ref{lbubble}) and (\ref{cosmicsource}).}}  \\
 \hline
A& None  & N/A & N/A &  N/A    \\
 B1  & Conduction  & 0.4 & N/A&N/A    \\
 B2 & Conduction  & 0.8 & N/A&N/A   \\
  B3 & Conduction  & 0.6 & N/A&N/A   \\
  C & Conduction, {CR\footnote{Cosmic-ray heating.}} & 0.3 & 0.003&  0.3   \\
  D1 & Conduction, CR & 0.1& 0.003  & 0.3    \\
 D2& Conduction, CR & 0.4& 0.003  & 0.3    \\
 E1 & Conduction, CR  & 0.3 & 0.05 & 0.3    \\
  E2 & Conduction, CR  & 0.3 & 0.0003 & 0.3   \\
 F1 & Conduction, CR  & 0.3 & 0.003 & 0.1   \\
  F2 & Conduction, CR  & 0.3 & 0.003 & 0.7   \\
  F3 & Conduction, CR  & 0.3 & 0.003 & 1.5   \\   
 {G1 \footnote{Runs G1, G2 and G3 are performed to check the dependence of our model on the CR diffusion coefficient. For run G1, $\kappa_{c}$ has the form of equation (\ref{diff_coef}) with $\kappa_{0}=3 \times 10^{27}$ cm$^{2}$ s$^{-1}$.}}
 & Conduction, CR  & 0.3 & 0.003 & 0.3   \\
  {G2 \footnote{For run G2, $\kappa_{c}$ has the form of equation (\ref{diff_coef}) with $\kappa_{0}=3 \times 10^{29}$ cm$^{2}$ s$^{-1}$.}}& Conduction, CR  & 0.3 & 0.003 & 0.3   \\
{G3 \footnote{For run G3, $\kappa_{c}$ is taken to be constant throughout the cluster: $\kappa_{c}=3 \times 10^{28}$ cm$^{2}$ s$^{-1}$.}}& Conduction, CR  & 0.3 & 0.003 & 0.3   \\     
 \hline
\label{table1}
\end{tabular}
\end{minipage}
\end{table}

 \begin{figure}
\includegraphics[width=0.45\textwidth]{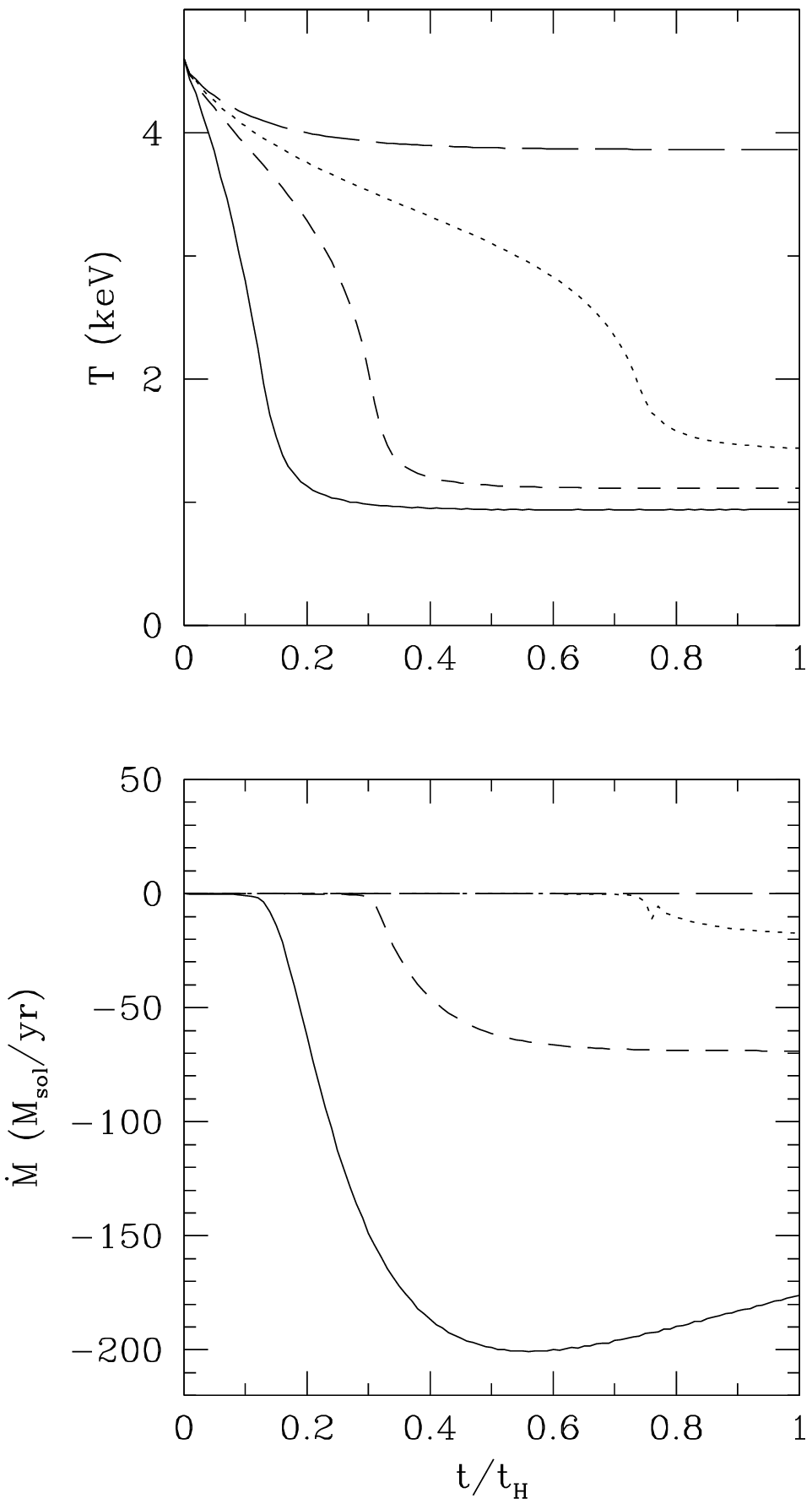}
\caption{Time evolution of gas temperature (\textit{upper panel}) and mass accretion rate (\textit{lower panel}) at $r=5$ kpc for runs A (\textit{solid line}), B1 (\textit{short-dashed line}), B2 (\textit{long-dashed line}), and B3 (\textit{dotted line}). See Table \ref{listtable} for additional information.}
 \label{plotsim0}
 \end{figure}
 
\begin{figure}
\includegraphics[width=0.45\textwidth]{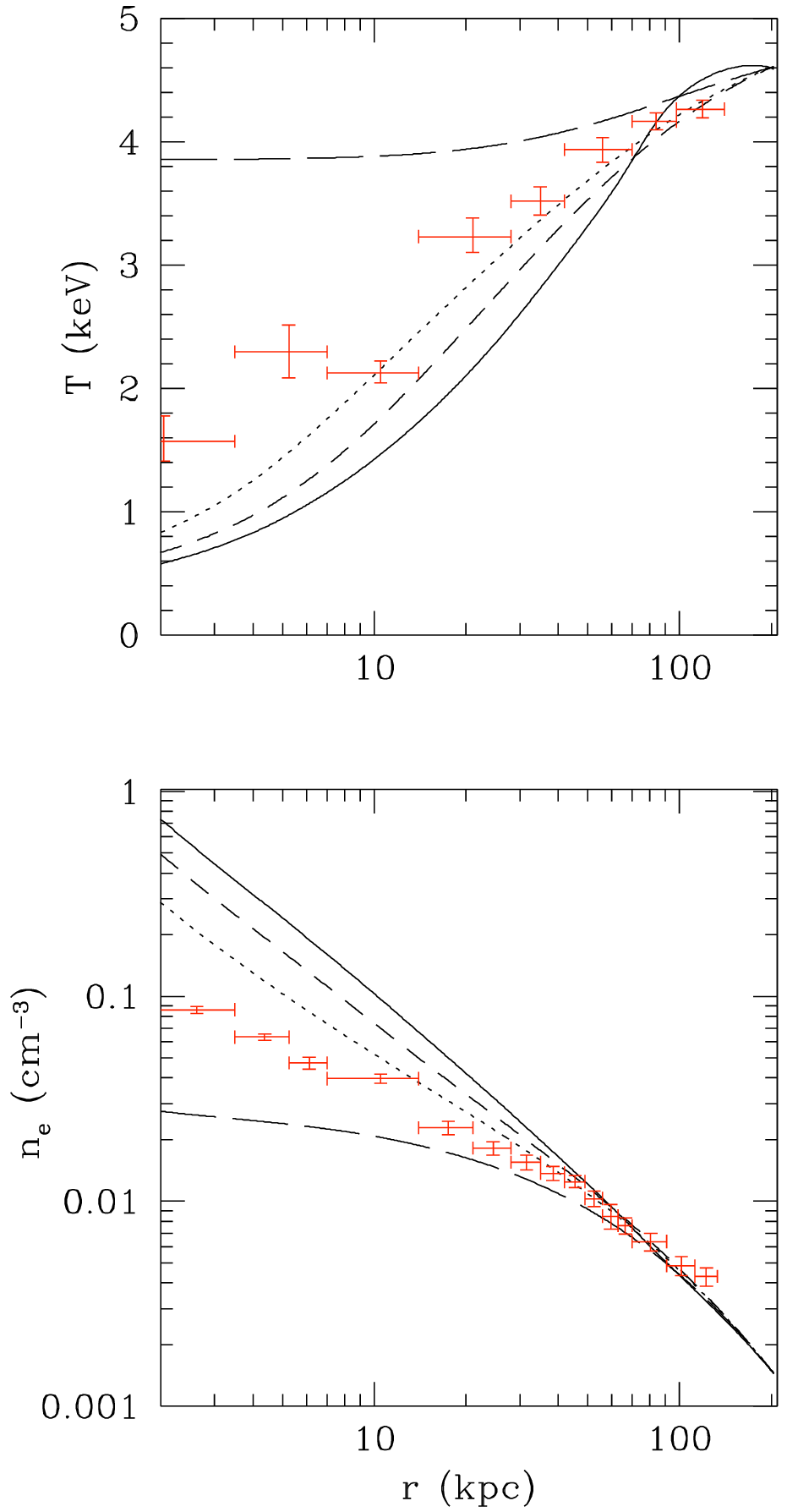}
\caption{Final density and temperature profiles after a Hubble time for runs A (\textit{solid line}), B1 (\textit{short-dashed line}), B2 (\textit{long-dashed line}), and B3 (\textit{dotted line}); crosses indicate {\it Chandra} data \citep{2002MNRAS.336..299J}. None of these models provide an adequate fit to the data.}
 \label{plotsim0b}
 \end{figure}

\section{Hydrodynamic Simulations}
\label{section:evolution}

In this section, we follow the long-term evolution of a series of spherically symmetric cluster models from a state far from equilibrium to investigate whether the cluster will relax to a stable quasi-equilibrium state. Basic information for the set of simulations presented in this paper is listed in Table \ref{listtable}. Our models are intended to be generic, but, for definiteness, we choose the cluster A2199 as our fiducial cluster. We compare the steady-state profiles of electron number density and temperature with the observational profiles as well.

\subsection{Simulation setup}
 
We use the ZEUS-3D hydrodynamic code \citep{1992ApJS...80..753S} in its one-dimensional mode; we gratefully acknowledge Mateusz Ruszkowski for supplying us with the modified version described in RB02, which includes radiative cooling and thermal conduction. We solve equations (\ref{cosmic1}), (\ref{hydro11}),~(\ref{hydro21}) and~(\ref{hydro31}) for our fiducial cluster Abell 2199; in particular, we have incorporated into ZEUS a background gravitational potential (eq. [\ref{gravpotent}]), cosmic-ray heating, cosmic-ray pressure support, cosmic-ray transport and the cosmic-ray energy equation (eq. [\ref{cosmic1}]).  For numerical stability, the conduction term is integrated using time steps that satisfy the Courant condition
\begin{equation}
\Delta t \leq \frac{1}{2}\frac{E_{\text{g}}(\Delta r)^{2}}{f \kappa_{\text{Sp}} T}
\text{.}
\end{equation}
The time steps in our simulation are also chosen to be small enough to satisfy the Courant conditions required by numerical stability of the cosmic-ray energy equation
\begin{equation}
\Delta t \leq \text{min}\left(\frac{\Delta r}{|u+v_{\text{A}}| }, \frac{(\Delta r)^{2}}{2 \kappa_{c}}\right)
\text{.}
\end{equation}

Our computational grid extends from $r_{\text{min}}=1$ kpc to $r_{\text{max}}=200$ kpc. In order to resolve adequately the inner regions, we adopt a logarithmically spaced grid in which $(\Delta r)_{i+1}/(\Delta r)_{i}=(r_{\text{max}}/r_{\text{min}})^{1/N}$, where $N$ is the number of active zones. 
We performed our main simulation (run C) in three different resolutions: $N=100$, $200$, $400$. The results of these three simulations are quite similar, with virtually identical results for the second two. Therefore, we are confident that our simulations are numerically convergent. The standard resolution of our simulations presented in this paper is $N=400$. 

For initial conditions, we assume the ICM to be isothermal at $T=4.6$ keV, and solve for hydrostatic equilibrium. We assume that at the outer boundary $r_{\text{max}}$, $n_{e}(r_{\rm max})=0.0015$ cm$^{-3}$, which is close to the value extrapolated from the observational density profile. We assume that the cosmic-ray energy density $E_{c}$ is a very small constant value throughout the cluster initially. Our results are not sensitive to this value, which is chosen to be $E_{c}=3.8 \times 10^{-14}$ ergs s$^{-1}$ cm$^{-3}$ in the models presented in the rest of the paper. For boundary conditions, we assume that the gas is in contact with a thermal bath of constant temperature and pressure at the outer radius, where the cooling time exceeds the Hubble time. Thus, we ensure that temperature and density of the thermal gas at the outer radius are constant.\footnote{These are the same boundary conditions used by RB02.} We extrapolate all hydrodynamic variables from the active zones to the ghost zones by allowing them to vary as a linear function of radius at both the inner and outer boundaries. The intracluster gas is allowed to flow in and out of active zones at both the inner and outer boundaries. Cosmic ray injection by the central AGN is only allowed when the gas at the inner boundary flows inward.

\begin{figure*}
\includegraphics[width=0.9\textwidth]{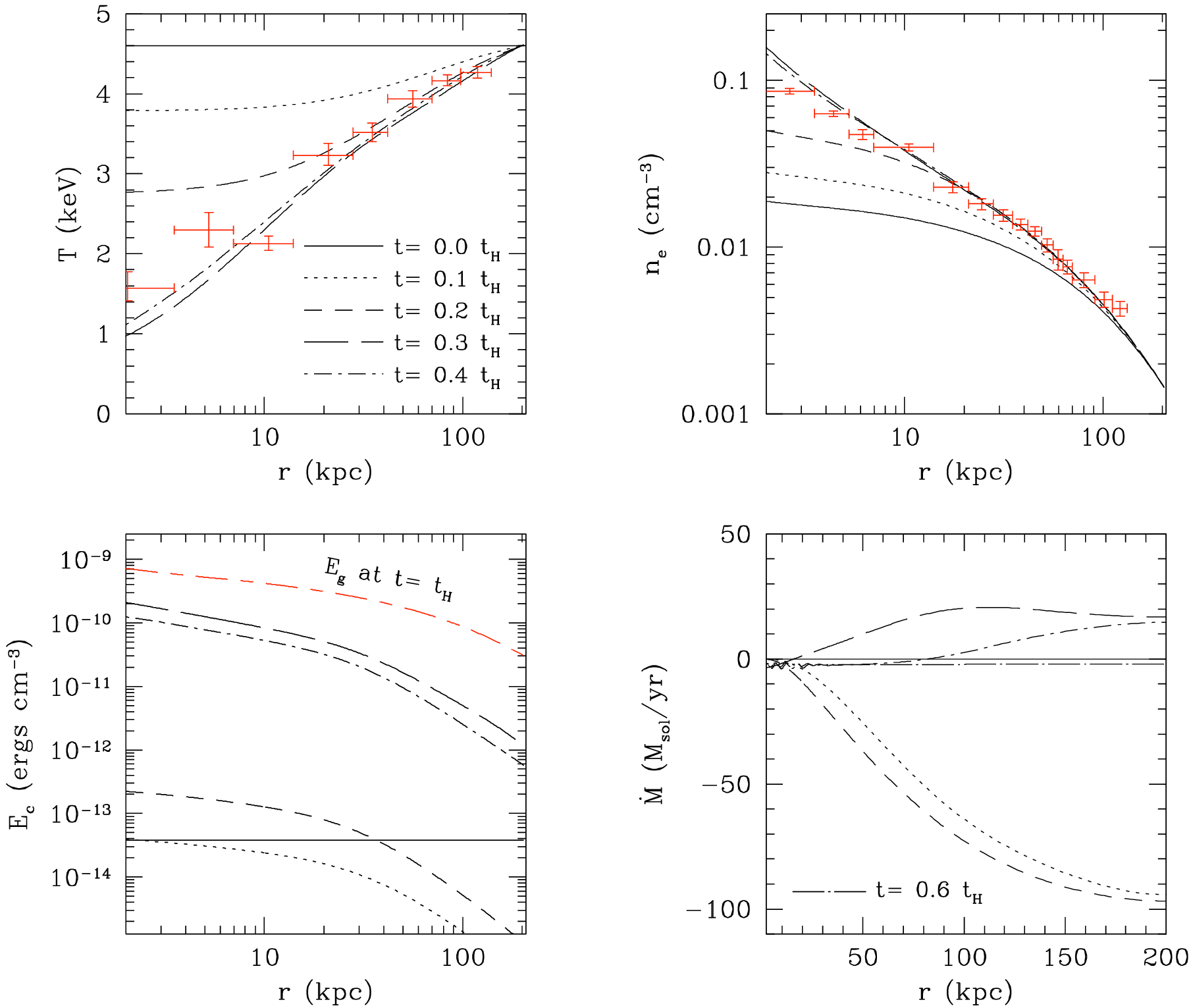}
\caption{Evolution of gas temperature (\textit{upper left}), electron number density (\textit{upper right}), cosmic-ray energy density (\textit{lower left}) and accretion rate plotted as a function of distance from the cluster center for run C. The cluster relaxes to a quasi-equilibrium state at $t \sim 0.4 t_{\text{H}}$, except that the mass accretion rate takes slightly longer to adjusts to steady state ($\dot{M}$ at $t=0.6t_{\text{H}}$ is plotted additionally to show its steady profile). Crosses in the upper panels indicate \textit{Chandra} data \citep{2002MNRAS.336..299J}. The short-long dashed curve in the lower left panel shows the final gas energy density.}
 \label{plotsim1}
 \end{figure*}

\begin{figure*}
\includegraphics[width=0.9\textwidth]{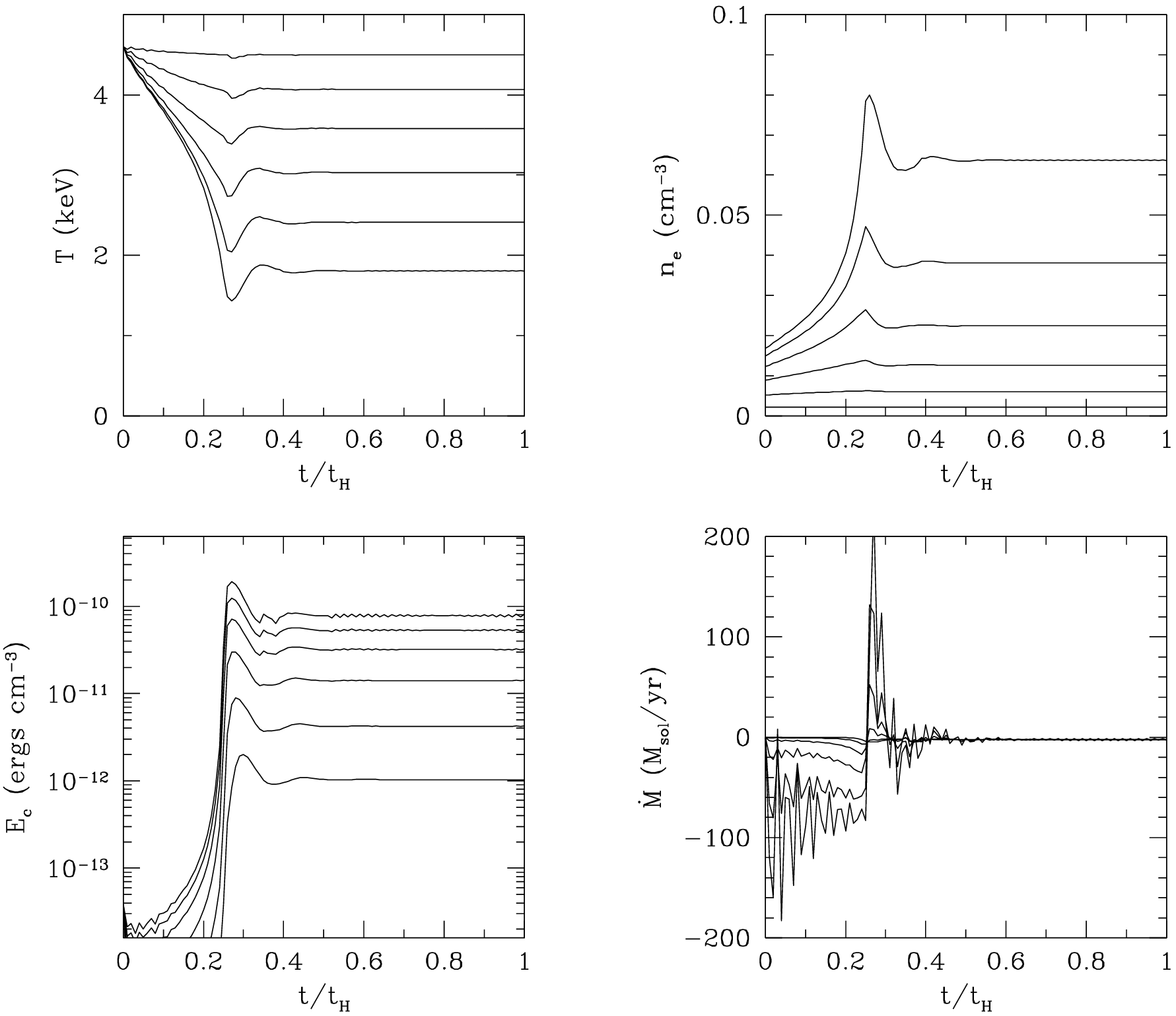}
\caption{Dependence of  gas temperature (\textit{upper left}), electron number density (\textit{upper right}), cosmic-ray energy density (\textit{lower left}) and accretion rate as a function of time for different distances from the cluster center for run C. See text for additional information.}
 \label{plotsim2}
 \end{figure*}

\subsection{The cooling flow model without any heating}

To establish a control, our first simulation (run A) follows the evolution of the ICM under the pure radiative cooling for a Hubble time $t_{\text{H}}=H_{0}^{-1}$ ($H_{0}=70$ km s$^{-1}$ Mpc$^{-1}$). The evolution is simple. Figure \ref{plotsim0} shows the time evolution of the gas temperature and mass accretion rate at $r=5$ kpc. As is clearly seen, the ICM cools catastrophically (characterized by a rapid decrease in the central temperature) and finally reaches a quasi-steady state, where a strong cooling flow ($\dot M \sim -200$ M$_{\sun}$ yr$^{-1}$) is formed. Note that the long-term evolution of the ICM with a strong cooling flow in our model may not be accurate: for instance, we neglect the deepening of the gravitational potential well due to the large mass deposition at the cluster center by the cooling flow \citep[see][]{1990ApJ...352..466M}. But this control establishes a minimal baseline for the amount of cooling expected if there are no heating sources.

\subsection{The model with conduction only}
\label{section:condsim}

In this subsection, we explore the role of thermal conduction in the evolution of the ICM. We first performed two simple simulations with $f=0.4$ (run B1) and $f=0.8$ (run B2), following the evolution of the ICM subject to radiative cooling and thermal conduction. The time evolution of the gas temperature and mass accretion rate at $r=5$ kpc is shown in Figure \ref{plotsim0}, while Figure \ref{plotsim0b} shows the final density and temperature profiles as a function of radius. Agreeing with \citet{1989ApJ...345..666G}, run B1 shows that a moderate level of thermal conduction delays the cooling catastrophe and reduces the mass accretion rate at the final quasi-steady state. However, weak conduction cannot suppress the cooling catastrophe sufficiently. Even with $f=0.4$, the final mass accretion rate at $r=5$ kpc is around $\dot M \sim -70$ M$_{\sun}$ yr$^{-1}$ and the final temperature at $r=5$ kpc is around $1.1$ keV, which is smaller than the observed value ($\sim 2$ keV). Interestingly, this value of $f=0.4$ is the valued needed to build an equilibrium model in which conduction balances cooling (e.g, \citet{2003ApJ...582..162Z}); we have verified this by building such a model which matches the observed temperature and density profiles for A2199, and found the eigenvalue $f=0.43$. If one perturbs around the equilibrium state, the global thermal instability of such models are also claimed to be dynamically unimportant \citep[the instability growth time is $\sim 2-5$ Gyr;][]{2003ApJ...596..889K}. However, if one starts far from equilibrium, then evolution toward the equilibrium profile is not guaranteed. We explore this and related issues in a forthcoming paper (Guo et al 2007, in preparation). On the other hand, strong conduction successfully prevents the cooling catastrophe, just as shown in Figure \ref{plotsim0}, but the temperature does not drop significantly toward the cluster center, in violation of the observed temperature gradient. 

Nonetheless, it may be possible to fine-tune the value of $f$ so that the final state (at $t \sim t_{\text{H}}$) of the pure conduction model produces a reasonably good fit to the observational data. We thus performed a pure conduction simulation with $f=0.6$ (run B3), which produced a somewhat better but still unsatisfactory fit to the data. With sufficient diligence it might be possible to find a satisfactory model, but the amount of fine-tuning seems excessive. Clearly, results depend sensitively on the assumed value of $f$: if $f$ is too large, the temperature profile is too close to isothermal; if it is too low, a strong cooling flow develops. Since neither nearly isothermal nor strong cooling flow clusters are observed, if only conduction balances cooling then $f$ must be restricted to a narrow range. Yet, the value of $f$ required to explain observed temperature and density profiles profiles differs from cluster to cluster, and a physical explanation of how $f$ self-adjusts in each cluster is missing; furthermore, a significant fraction of observed clusters cannot be fit at all by conduction only models with $f \le 1$ \citep{2003ApJ...582..162Z}. By contrast, we find in \S \ref{section:evolution2} that if a secondary heating mechanism such as cosmic-ray heating is included, this fine-tuning problem is eliminated, and a broad range of $f$ is permissible, with the remainder of the heating being supplied by CR heating in a self-regulating fashion. Note also from Figure \ref{plotsim0} that the ICM in run B3 takes a long time (comparable to the Hubble time) to reach a steady state. As we shall see in the next subsection, by including a physically motivated feedback heating term, the ICM will relax more quickly to a steady state, which produces a much better fit to the observational data as well.
 
Our conduction model (eq. [\ref{hydro3}]) in this paper is somewhat idealized. In reality, both electron conduction and turbulent mixing may contribute to heat transport in clusters \citep[for a comprehensive discussion, see][]{2006ApJ...645L..25L}. Note that the turbulent mixing model will probably suffer a similar fine-tuning problem (fine-tuning of the mixing parameter may be required, see \citet{2003ApJ...596L.139K}).

\subsection{The model with conduction and cosmic-ray feedback heating}
\label{section:evolution2}

\begin{figure}
\includegraphics[width=0.45\textwidth]{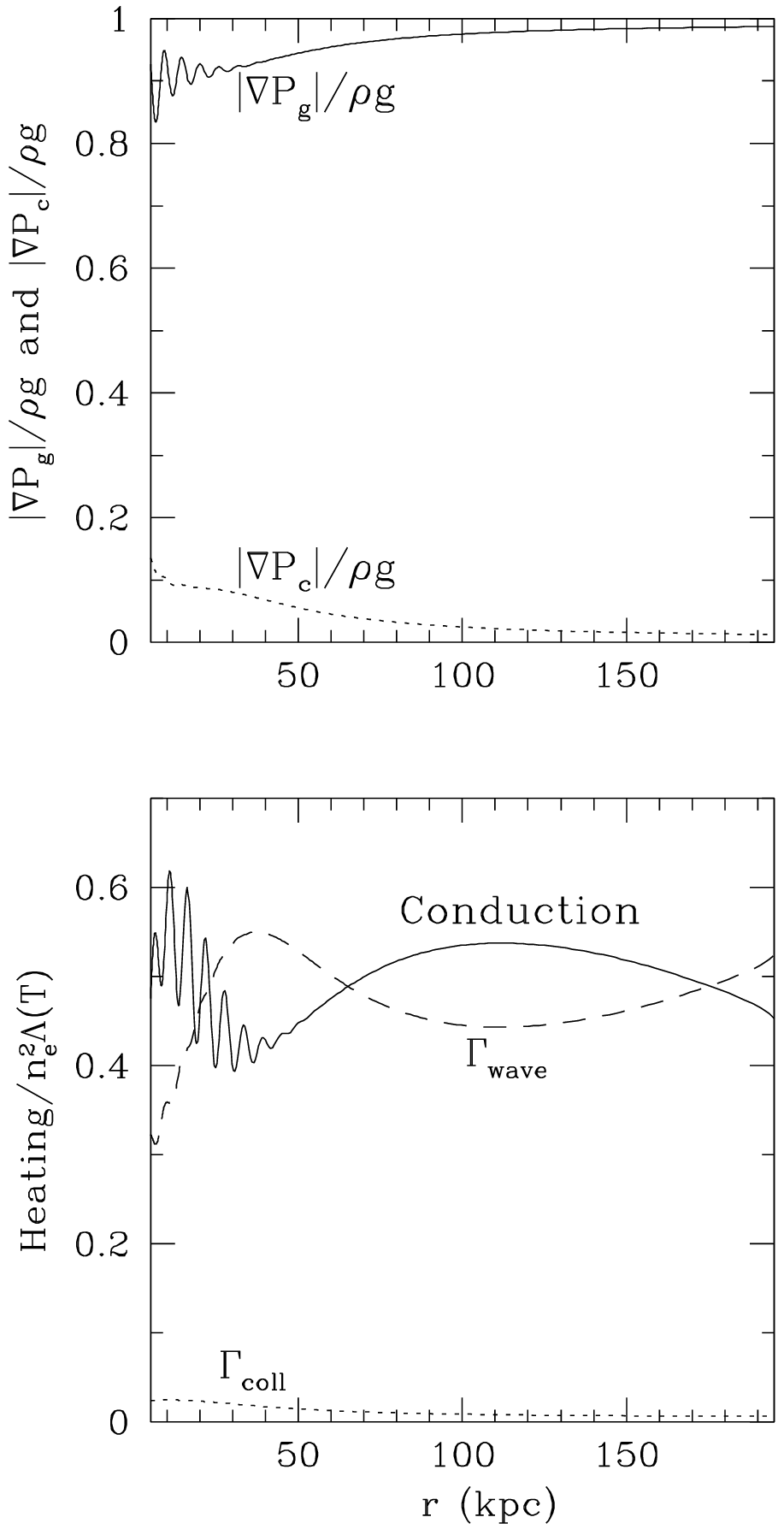}
\caption{Radial profiles of pressure support (\textit{upper panel}) and relative importance of various heating mechanisms (\textit{lower panel}) in the final state of our main simulation presented in \S \ref{section:evolution2} (run C). Note that the short-wavelength oscillations in the curves are caused by sound waves due to changing boundary conditions. See text for additional information.}
 \label{plotsim3}
 \end{figure}
 
The main results of this paper are presented in this subsection. We consider the evolution of the ICM subject to radiative cooling, thermal conduction and feedback heating by cosmic rays. To illustrate our results, we present one representative model (run C) with the following parameters: $f=0.3$, $\epsilon=0.003$, $\nu =0.3$. We follow cluster evolution for a Hubble time $t_{\text{H}}$. Figure \ref{plotsim1} shows the radial profiles of gas temperature, electron number density, cosmic-ray energy density and mass accretion rate for a few time epochs. This model settles down to a stable steady state at $t \sim 0.4 t_{\text{H}}$. Figure \ref{plotsim2} shows the gas temperature, electron number density, cosmic-ray energy density and mass accretion rate as a function of time (in units of the Hubble time) for different distances from the cluster center $r=5$, $10$, $20$, $40$, $80$, $160$ kpc. In the case of electron number density and cosmic-ray energy density, the above sequence of $r$ corresponds to the curves from top to bottom. For the gas temperature, the trend is the opposite. In the case of accretion rate, the amplitude of oscillations increases with $r$. These oscillations are caused by sound waves, which propagate across the cluster as it adjusts to changing boundary conditions (see RB02). The precise character of these sound waves depends on the resolution and boundary conditions. After the cluster relaxes to the quasi-steady state, similar numerical oscillations with a small amplitude and a short wavelength also appear near the cluster center, as is readily seen in the time evolution curves of cosmic ray energy density and mass accretion rate in Figure \ref{plotsim2}, as well as spatially in the pressure gradient and heating terms in Figure \ref{plotsim3}. 

The evolution of the cluster is very similar to that of RB02, since their AGN feedback heating is also triggered by the mass accretion. Strong X-ray emission in the cluster center leads to a gradual decrease in temperature and a slow increase in gas density, which in turn increases the cooling and thus increases the accretion rate. Cosmic ray injection is controlled by the mass accretion rate at the cluster center. As the CR injection rate increases, the CR heating rate also increases and hence, the cluster does not cool in a runaway fashion. As can be clearly seen in Figure \ref{plotsim2}, the slow evolution of the cluster is followed by a cooling catastrophe at $t\sim 0.25t_{\text{H}}$. Unlike the standard cooling flow models where the gas cools to very low temperatures and a strong cooling flow forms, the feedback heating mechanism in our model suppresses the cooling catastrophe quickly after its onset. The gas temperature, electron number density and cosmic-ray energy density then cycle up and down for several times when the cluster adjusts its mass accretion rate in response to the relative importance of heating and cooling, and are stabilized within $\sim 0.1t_{\text{H}}$. The cluster thus relaxes to a stable steady state. The evolution of mass accretion rate is similar. After oscillations through positive and negative values, the mass accretion rate tends to a small constant negative value. This value is also approximately constant at all radii, as it should be for a steady state cluster. In the final state, the mass accretion rate is about $\dot M \sim -2.3$ M$_{\sun}$ yr$^{-1}$, which is much smaller than accretion rates inferred from the standard cooling flow models, and consistent with the observed upper bounds of $\dot M \lsim 12 {\rm M_{\odot} \, yr^{-1}}$ \citep{2002MNRAS.336..299J}.
 
Thus, a cooling catastrophe is averted in our model. Starting from a state far from equilibrium, the cluster relaxes to a sustainable and stable steady quasi-equilibrium state. In the upper panels of Figure \ref{plotsim1}, the observational temperature and electron density profiles \citep{2002MNRAS.336..299J} are also shown. As is clearly seen, the final steady state of our model produces a very good fit to the observational profiles. In the lower left panel of Figure \ref{plotsim1}, we also show the radial profile of thermal energy density in the final state. In steady state, the ratio of cosmic-ray pressure to thermal pressure ($P_{\text{c}}/P_{\text{g}}$) is always less than $0.1$ and decreases away from the cluster center. This is well within upper bounds in nearby rich clusters of $P_{\text{c}}/P_{\text{g}} \lsim 20\%$ (\citet{1997ApJ...477..560E}, Virgo and Perseus clusters) and $P_{\text{c}}/P_{\text{g}} \lsim 30\%$ (\citet{2004A&A...413...17P}, Coma cluster). 

The upper panel of Figure \ref{plotsim3} shows the steady-state ratios of thermal and CR pressure gradients to the gravity. As is readily seen, the thermal pressure support dominates over the whole cluster, although cosmic rays provide a small amount of pressure support in the central regions of the cluster ($\sim 0.1 \rho g$ in the central $\sim 30$ kpc). In the lower panel of Figure \ref{plotsim3}, we show the relative importance of various heating mechanisms in steady state. The heating due to thermal conduction is comparable to the cosmic ray heating, which is dominated by wave heating through the dissipation of the hydromagnetic waves, while cosmic-ray collisional heating is negligible. The volume-integrated cosmic-ray heating rate amounts to $1.1\times 10^{44}$ ergs s$^{-1}$ in the final state, while the final X-ray luminosity is $2.3 \times 10^{44}$ ergs s$^{-1}$. The volume-integrated cosmic-ray injection rate in the final steady state is $\sim 2.4\times 10^{44}$ ergs s$^{-1}$, where around half of the cosmic-ray energy is transported to outer regions of the cluster ($>200$ kpc) and heats the ICM in those regions. Only a small fraction ($\sim 2\%$) of the CR energy escapes the cluster in the form of gamma rays and neutrinos. 
 
\begin{figure*}
\includegraphics[width=0.9\textwidth]{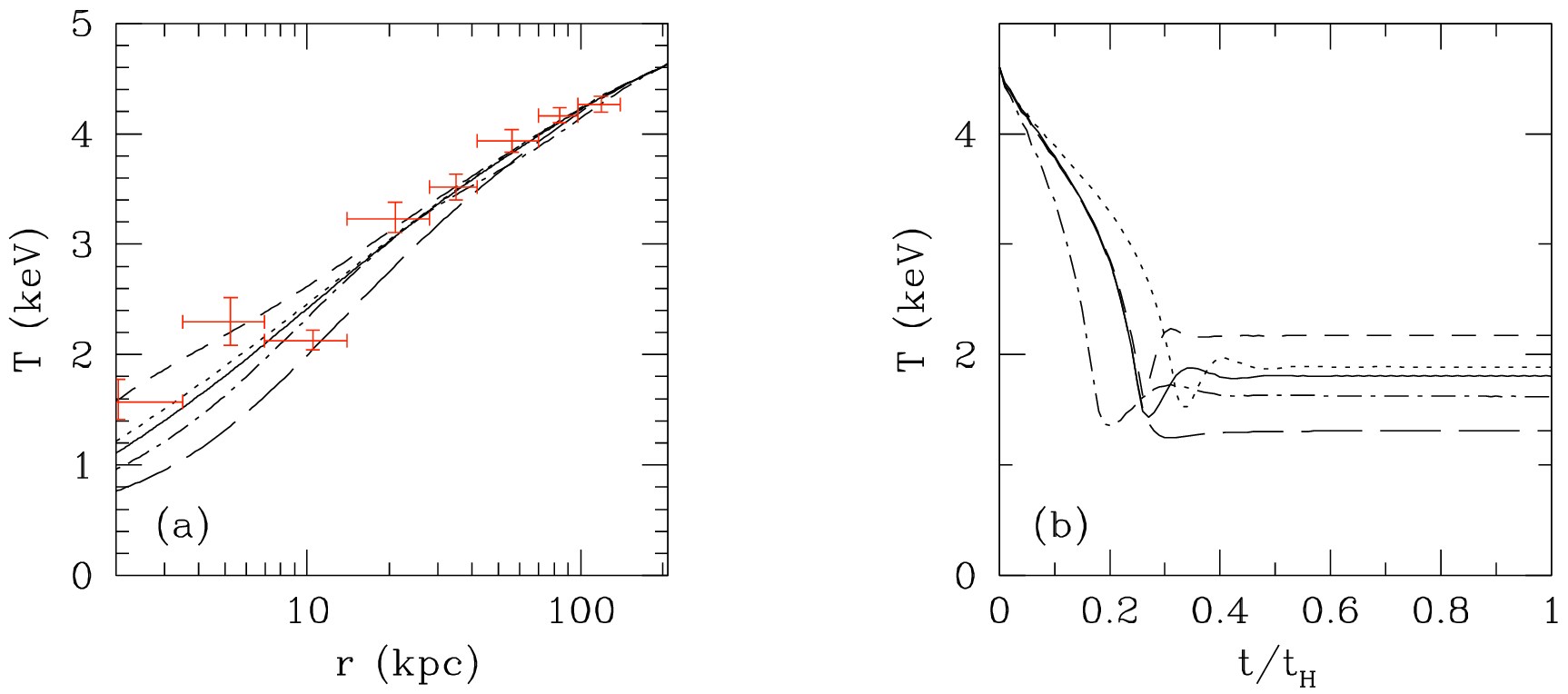}
\caption{(a) Radial profile and (b) time evolution of gas temperature at $r=5$ kpc for runs C (\textit{solid line}), D1 (\textit{dot-dashed line}), D2 (\textit{dotted line}), E1 (\textit{short-dashed line}), and E2 (\textit{long-dashed line}). See Table \ref{listtable} for additional information.}
 \label{plotsim4}
 \end{figure*}
 
Our models and results are very robust to the level of thermal conduction. Runs D1 and D2 follow the evolution of the cluster with a lower thermal conductivity ($f=0.1$) and a higher thermal conductivity ($f=0.4$), respectively. The steady-state profiles of gas temperature and electron number density for both runs are very close to those for run C ($f=0.3$). The temperature profiles in the final state for both runs are shown in Figure \ref{plotsim4}a. Figures \ref{plotsim4}b shows the time evolution of gas temperature at $r=5$ kpc. As is clearly seen, thermal conduction delays the onset of the cooling catastrophe and, with a higher level of thermal conduction, the cluster approaches the steady state later, which agrees with similar results from the pure conduction models (see \S~\ref{section:condsim}).

We also run similar simulations with different efficiency of feedback (run E1 with $\epsilon=0.05$ and run E2 with $\epsilon=0.0003$). As shown in Figure \ref{plotsim4}b, the cooling catastrophe happens at almost the same time ($\sim 0.25 t_{\text{H}}$). However, with higher efficiency, the cooling catastrophe is more strongly suppressed and the final mass accretion rate is more reduced. For the very low efficiency $\epsilon=0.0003$, the cooling catastrophe is less suppressed and the gas at inner radii can cool to lower temperatures and higher densities in the final steady state, as clearly seen in Figure \ref{plotsim4}a. The final mass accretion rate in this low-efficiency model is about $\dot M \sim -19$ M$_{\sun}$ yr$^{-1}$, which is still much smaller than accretion rates in standard cooling flow models, and marginally consistent with the rough observational bound of $\dot M \lsim 12 {\rm M_{\odot} \, yr^{-1}}$ \citep{2002MNRAS.336..299J}.

\begin{figure}
\includegraphics[width=0.45\textwidth]{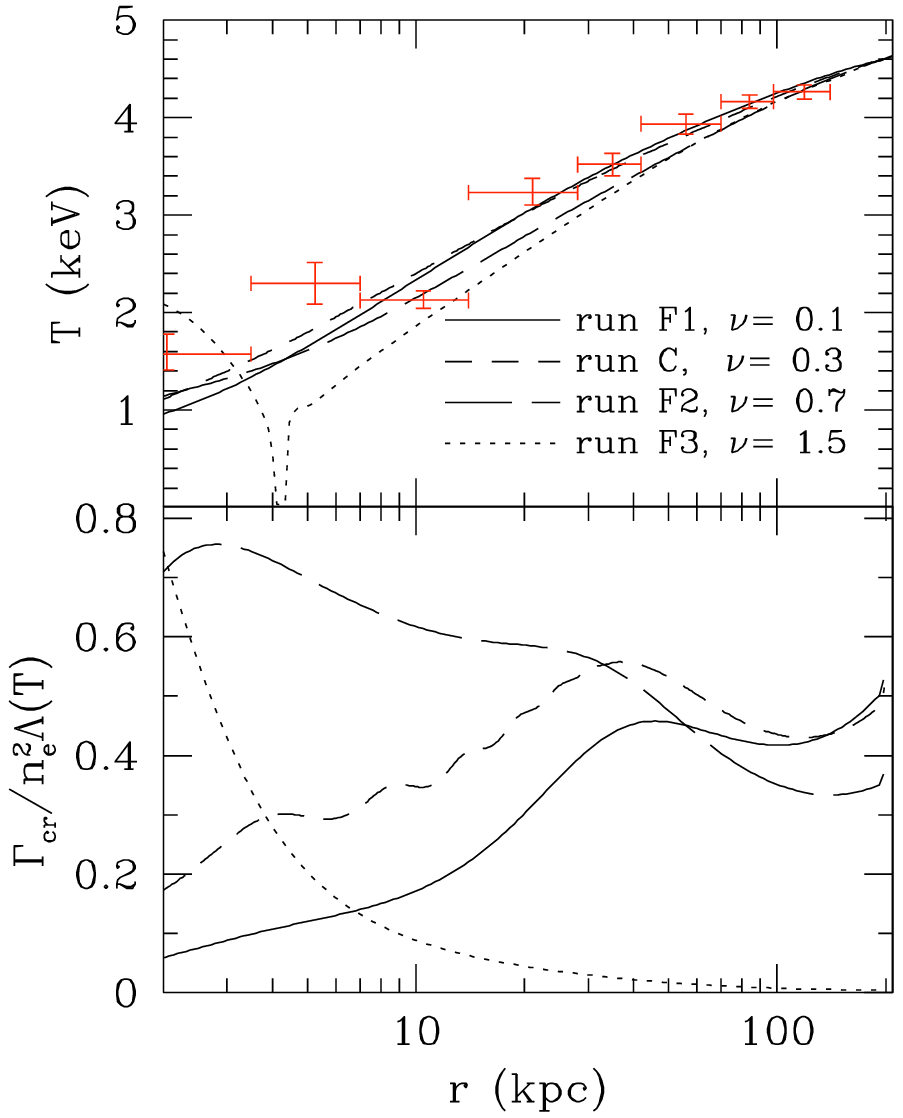}
\caption{Radial profiles of gas temperature (\textit{upper panel}) and ratio of cosmic-ray heating rate to gas cooling rate (\textit{lower panel}) for runs C, F1, F2, and F3. The curves for runs C, F1 and F2 are plotted at $t=0.4t_{\text{H}}$, when the cluster has relaxed to steady state. For run F3, the curve in the lower panel is plotted at $t=0.25t_{\text{H}}$, which shows clearly that the cosmic-ray heating is centrally peaked, resulting in insufficient heating in the outer regions of the cluster and thus leading to a cooling catastrophe at $r\sim 4$ kpc as shown in the temperature profile (\textit{upper panel}) plotted at $t=0.3t_{\text{H}}$.}
 \label{plotsim5}
 \end{figure}

Our models are also very robust to the value of $\nu$, which determines the spatial distribution of cosmic-ray injection into the ICM. Runs F1 and F2 follow the evolution of the cluster with a lower value ($\nu=0.1$) and a higher value ($\nu=0.7$) of $\nu$, respectively. The cluster relaxes to steady state at almost the same time as run C ($\nu=0.3$); the steady-state profiles of gas temperature and density are also very similar to those for run C. Figure \ref{plotsim5} shows the steady-state radial profiles of gas temperature and ratio of cosmic-ray heating rate ($\Gamma_{\text{cr}}=\Gamma_{\text{wave}}+\Gamma_{\text{coll}}$) to gas cooling rate for these runs. As is readily seen, for higher values of $\nu$, the cosmic ray injection is more centrally localized, and so is the cosmic ray heating. The resulting steady state mass accretion rate decreases with $\nu$ ($\dot{M} \sim -6.6$, $-2.3$ and $-1.0$ M$_{\sun}$ yr$^{-1}$ for runs F1, C and F2, respectively). With a combination of cosmic ray heating and thermal conduction, our model produces a reasonably good fit to observation for a broad range of $\nu$. Note that for $\nu \lsim 0.1$, the PdV work during the bubble expansion dominates over the CR injection (see Fig. \ref{pdvcr}) and thus becomes the main mechanism transferring the AGN mechanical energy into the thermal energy of the ICM. We also run a simulation (run F3) with a very high value of $\nu$ ($\nu=1.5$), where the cosmic rays are essentially dispersed into the ICM at the cluster center. As shown in Figure \ref{plotsim5}, the cosmic rays are unable to directly heat the ICM at large radii and the cluster suffers a cooling catastrophe at  $r\sim 4$ kpc at $t \sim 0.3t_{\text{H}}$. The appropriate value of $\nu$ depends on the bubble disruption rate and is fairly uncertain (see discussion in \S \ref{section:AGN_injection}). However, a very high value of $\nu$ seems unlikely since bubbles are observed to survive out to large projected radii. 

Recent studies \citep{2005ApJ...632..809C, 2006ApJ...642..140C} suggest that the cosmic ray pressure gradient may drive convection, if the convective instability criterion 
\begin{equation}
\frac{\mu_{\text{e}}}{\mu}n_{\text{e}}k_{\text{B}}\frac{dT}{dr}<-\frac{dP_{\text{c}}}{dr} \label{mti}
\end{equation}
is satisfied. In cluster cores, a strong negative cosmic ray pressure gradient is required to drive convection, since the gas temperature increases away from the cluster center. When $\nu$ is higher, the distribution of cosmic ray injection is more centrally peaked, and thus the resulting cosmic-ray pressure gradient is more negative. We checked the convective instability criterion (eq. [\ref{mti}]) for runs F1 ($\nu=0.1$), C ($\nu=0.3$), and F2 ($\nu=0.7$), and found that the cluster is always convectively stable during the simulations. In the steady state configuration of our main model (run C), we find that the ratio of the left-hand side to the right-hand side of equation (\ref{mti}) is $\sim 5-10$. For run F3 ($\nu=1.5$), where cosmic rays are mainly dispersed into the ICM at the cluster center, the instability criterion (eq. [\ref{mti}]) is easily met at the cluster center at the very beginning of the simulation, suggesting that the cluster becomes convectively unstable long before the onset of the cooling catastrophe. Convection driven by the cosmic-ray pressure gradient thus provides an alternative means for heating the ICM and generating the needed magnetic turbulence in cluster cores \citep{2005ApJ...632..809C}, which is obviously beyond the scope of this paper. 
 
 \begin{figure}
\includegraphics[width=0.45\textwidth]{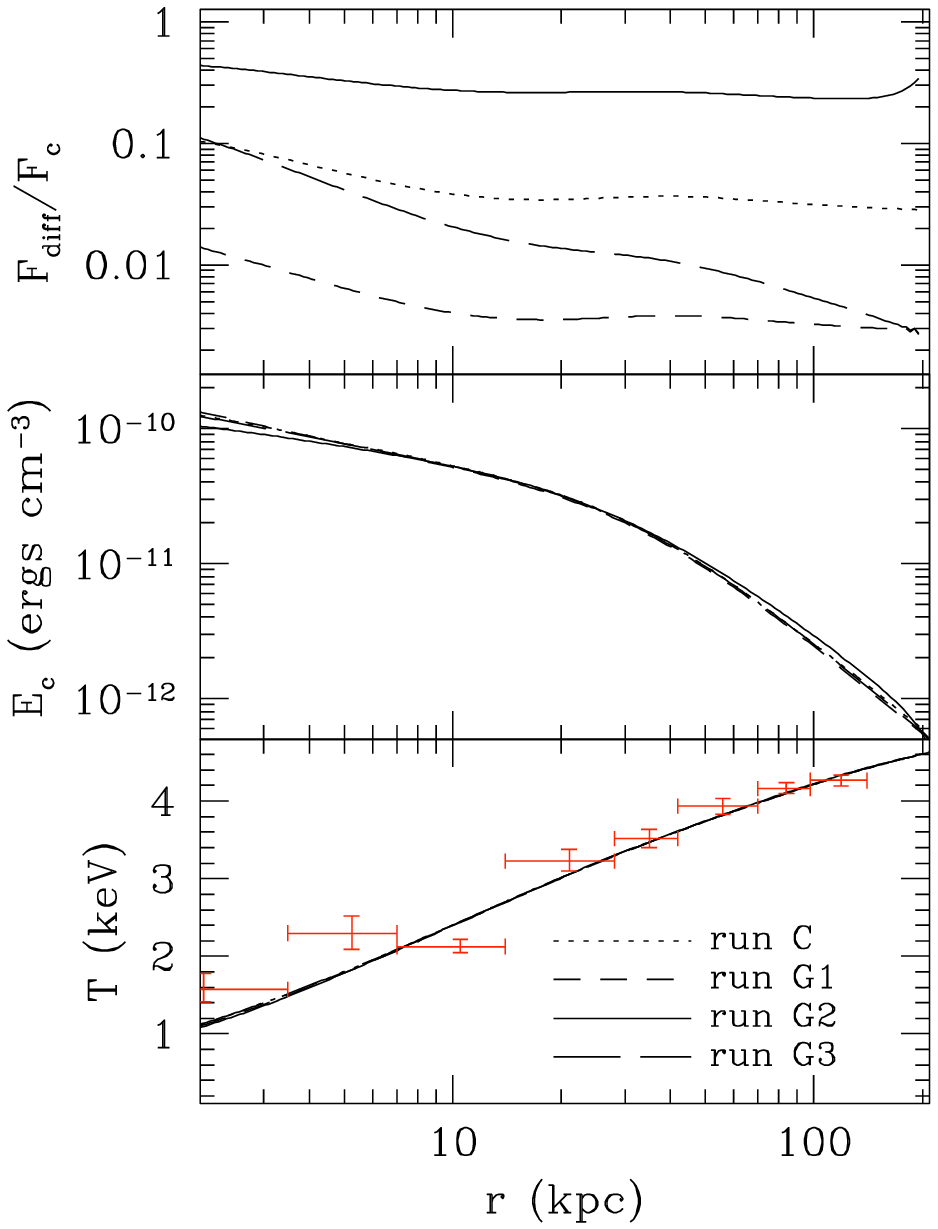}
\caption{Radial profiles of the contribution of cosmic-ray diffusion to CR energy flux (\textit{upper panel}), cosmic-ray energy density (\textit{middle panel}), and gas temperature (\textit{lower panel}) for runs C, G1, G2, and G3 at $t=0.4t_{\text{H}}$, when the cluster has relaxed to steady state. Note that the cluster relaxes to almost the same steady-state temperature profiles at almost the same time for these runs, which assume different radial profiles for $\kappa_{\text{c}}$. See Table \ref{listtable} for additional information.}
 \label{plotsim6}
 \end{figure}
 
The CR diffusion coefficient, $\kappa_{c}$, in galaxy clusters is fairly unclear. To check the dependence of our model on it, we performed three additional simulations with different radial profiles of $\kappa_{c}$, which has the form of equation (\ref{diff_coef}) with $\kappa_{0}=3 \times 10^{27}$ cm$^{2}$ s$^{-1}$ and $3 \times 10^{29}$ cm$^{2}$ s$^{-1}$ for runs G1 and G2 respectively. For run G3, $\kappa_{c}$ is taken to be constant throughout the cluster: $\kappa_{c}=3 \times 10^{28}$ cm$^{2}$ s$^{-1}$. The steady-state profiles of the contribution of cosmic ray diffusion, $F_{\text{diff}}=-\kappa_{\text{c}}(\partial E_{\text{c}}/\partial r$), to CR transport for these runs are very different, as shown in the upper panel of Figure \ref{plotsim6}. The middle and lower panels show the steady-state profiles of CR energy density and gas temperature. These runs demonstrate that the evolution of the ICM is very insensitive to the radial profiles of $\kappa_{c}$. This is due to the fact that the cosmic-ray diffusion time is much longer than the gas cooling time (see \S~\ref{section:AGN_injection}). In our model, the distribution of the cosmic rays in the cluster is mainly determined by the spatial distribution of CR injection into the ICM from the rising bubbles (see \S~\ref{section:AGN_injection}). Note that the diffusion is also usually subdominant with respect to the cosmic ray advection. The steady-state ratio of diffusion to advection for run C is  $\lsim 0.1$ in the central $\sim 10$ kpc and becomes negligible ($\sim 0.03$) in the outer regions; run G2 is an extreme case where the CR diffusion becomes comparable to the advection (see Figure \ref{plotsim6}).
 
\section{Conclusions}
\label{section:conclusion}

Awareness of the significant role cosmic rays could play in shaping the thermal and dynamical state of gas in galaxy clusters has been growing in recent years. Observations of diffuse radio synchrotron radiation from galaxy clusters imply that strong sources of non-thermal particles are indeed present. At the same time, recent studies show that active galactic nuclei inflate buoyant bubbles containing non-thermal radio emitting particles and could potentially play a central role in suppressing the cooling flows in cool core clusters. Many studies have focused on the potential dynamical effects of cosmic ray pressure support, but none have built successful models in which cosmic-ray heating is signficant. In this paper, we propose a new model of AGN feedback heating, in which cosmic rays produced by accretion-triggered AGN activity heat the ICM efficiently, with only a small dynamical perturbation on the ICM. 

% see the paragraph below
In our model, the cosmic rays are injected into the ICM mainly from the rising bubbles generated by central AGN activity, which is treated in a time-averaged sense. We assume that the cosmic rays are injected into the ICM instantaneously and neglect any time delay between central AGN activity and the cosmic ray injection. Such time-averaging is justifiable because the AGN duty cycle is much shorter than the gas cooling time. The cosmic rays then stream along the magnetic field lines in the ICM. Due to the cosmic ray streaming instability, Alfv$\acute{\text{e}}$n waves propagating nearly in the direction of the CR streaming are excited and scatter the cosmic rays in pitch angle. These waves grow exponentially until dissipated by nonlinear Landau damping, and thus heat the ICM efficiently. We note that the cosmic ray streaming may also depend on the details of the CR scattering by the small-scale MHD turbulence in the ICM, which is still poorly understood (see  \S~\ref{section:hydrowave}). Here only Alfv$\acute{\text{e}}$n waves self-excited by the cosmic-ray streaming instability are considered. 

We have performed a set of one-dimensional numerical simulations of the ICM, which is subject to radiative cooling, thermal conduction and cosmic ray heating. If only thermal conduction operates, extreme fine-tuning of the conduction suppression factor $f$ is required: if $f$ is too low, then a strong cooling flow develops. If $f$ is too high, the temperature profile becomes nearly isothermal, in contrast to observations where the temperature invariably declines toward the cluster center. On the other hand, once cosmic ray heating is including, our results are very robust to the level of thermal conduction: the reduced cooling flow in our new model automatically adjusts itself to some low value of the mass accretion rate, which is mainly determined by the value of efficiency $\epsilon$ in equation (\ref{lbubble}). Furthermore, unlike the conduction-only case, the conduction+CR heating case rapidly equilibrates toward a steady-state solution. For a representative model (run C), the ICM relaxes to a stable quasi-equilibrium state which is a very good fit to the observed gas temperature and density profiles. The cosmic-ray pressure in steady state is much less than the gas pressure ($P_{\text{c}} \lesssim 0.1 P_{\text{g}}$), while $\nabla P_{c} \lsim 0.1 \rho g$ in the central $\sim 30$ kpc and becomes negligible in the outer regions, all well within observational constraints. 

Thus, cosmic-ray heating models are a very attractive alternative to mechanical heating models \citep[e.g.][]{2002Natur.418..301B, 2004ApJ...611..158R, 2006ApJ...645...83V, 2005MNRAS.357..242R} in which the ICM is heated by the $pdV$ work of the expanding bubbles, viscous dissipation of emitted sound waves or mixing of the hot bubble plasma with the ICM. The detailed microphysics of how the latter processes take place has not been hammered out in detail, leaving a good deal of uncertainty; a definitive explanation for how energy is transported from the observed bubbles to the ICM in a distributed and isotropic fashion is still outstanding. Which is not to say that the cosmic-ray heating model presented here does not suffer from similar uncertainties: the details of how cosmic rays leak from the bubbles, and/or the rate at which bubbles are disrupted, are all highly uncertain. Nonetheless, the results presented here suggest that more detailed studies of cosmic-ray heating in fully 3D, cosmological simulations (e.g., the simulations of \citet{jubelgas06,pfrommer06} where most of the relevant cosmic-ray physics is already included) are warranted. At the same time, elucidating the details of bubble disruption/cosmic-ray diffusion would be very useful in determining whether cosmic-rays or mechanical processses provide a more efficient means of transporting heat from the bubble to the ICM. 

All of these issues may assume great urgency if {\it GLAST} detects the $\gamma$-ray signature from the decay of neutral pions produced when cosmic rays collide with ICM nucleons. For the particular model of A2199 (redshift $z=0.0309$; \citet{2002MNRAS.336..299J}) presented here (run C), and assuming that $\sim 1/3$ of hadronic losses go toward producing neutral pions and hence gamma-rays, we find a steady-state gamma-ray flux of $\sim 9.4 \times 10^{-13} {\rm erg \, s^{-1} \, cm^{-2}}$, which varies up to a factor of $\sim 2-3$ for other runs, depending on the model parameters, and should be a $\sim 3 \sigma$ detection for {\it GLAST}. Since AGN activities in real clusters are likely episodic, we note that this value may be viewed as a time-averaged estimate and the real $\gamma$-ray flux may be somewhat different, depending on the AGN duty cycle and the cosmic-ray injection rate (the maximum value of $\gamma$-ray flux during the cluster evolution in our main simulation (run C) is around twice greater than that in the final steady state). For fainter clusters, it should be possible to stack signals to provide a population-averaged limit \citep{2007arXiv0705.2588A}. 

\section*{Acknowledgments}

We are grateful to Mateusz Ruszkowski for providing us with his modified ZEUS code for their mechanical AGN heating models (RB02). We thank Shane Davis, Mateusz Ruszkowski and Mark Voit for discussions and comments on the manuscript, and the anonymous referee for a detailed and helpful report. We also thank Philip Chang and Evan Scannapieco for helpful discussions, and Torsten A. En\ss lin, Christoph Pfrommer and Sergio Colafrancesco for useful email correspondence. This work was supported by NASA grant NNG06GH95G and NSF grant AST-0407084. 

\bibliography{ms} 

\appendix

\section[]{The Time-dependent Cosmic-ray Equations}

In this Appendix, we derive the time-dependent energy equation for the cosmic rays in a background magnetized plasma (also see \citet{1982A&A...116..191M}; \citet{1991A&A...251..713K}). When the cosmic rays are streaming along the magnetic field lines at a speed faster than the local Alfv$\acute{\text{e}}$n speed, they excite hydromagnetic waves (e.g., Alfv$\acute{\text{e}}$n waves) by the cosmic-ray streaming instability \citep{1967ApJ...147..689L, 1969ApJ...156..445K}. The cosmic rays are then scattered in pitch angle and confined by these waves \citep[see][]{1971ApJ...170..265S}, for which we only consider forward Alfv$\acute{\text{e}}$n waves propagating nearly parallel to the background magnetic field in the direction of the local cosmic-ray streaming (backward Alfv$\acute{\text{e}}$n waves are damped; see \citet{2005ppfa.book.....K}, \citet{1967ApJ...147..689L}, \citet{1969ApJ...156..445K}). Including a net particle source function $Q$ (other than compression or expansion), the cosmic-ray transport equation may be written as \citep[e.g.][]{1971ApJ...170..265S, 1975MNRAS.172..557S}
\begin{align}
\frac{\partial f_{p}}{\partial t}+(\boldsymbol{u}+\boldsymbol{v}_{\text{A}})\boldsymbol{\cdot} \boldsymbol{\nabla} f_{p}=&\boldsymbol{\nabla} \boldsymbol{\cdot} ( \kappa_{\text{p}}  \boldsymbol{n}\boldsymbol{n} \boldsymbol{\cdot \nabla} f_{p}) \notag \\
&+\frac{1}{3}p
\frac{\partial f_{p}}{\partial p}  \boldsymbol{\nabla \cdot} (\boldsymbol{u}+\boldsymbol{v}_{\text{A}}) +Q
\text{,} \label{skilling}
 \end{align}
where $f_{p}(\boldsymbol{x}, p, t)$ is the isotropic part (in momentum) of the cosmic-ray phase space distribution function, $\boldsymbol{\nabla}\equiv \partial / \partial \boldsymbol{x}$, $\boldsymbol{u}$ is the velocity of the background plasma, $\boldsymbol{v}_{\text{A}}$ is the local Alfv$\acute{\text{e}}$n velocity, $\boldsymbol{n}=\boldsymbol{v}_{\text{A}}/v_{\text{A}}$ is a unit vector along the local magnetic field, and $\kappa_{\text{p}} $ is the diffusion coefficient given by
\begin{equation}
\kappa_{\text{p}} (\boldsymbol{x}, p)=v^{2}\left<\frac{1-\mu^{2}}{2\nu(\boldsymbol{x}, p, \mu)}\right>=\frac{v^{2}}{4}\int_{-1}^{+1}\frac{1-\mu^{2}}{\nu(\boldsymbol{x}, p, \mu)}d\mu \text{,}
\end{equation}
where $v=p[1+p^{2}/(m^{2}c^{2})]^{-1/2}/m$ is the cosmic-ray particle speed, $\mu=\boldsymbol{p \cdot n}/p$ is the pitch-angle cosine and $\nu(\boldsymbol{x}, p, \mu)$ is the frequency of pitch angle scattering of cosmic rays by hydromagnetic waves.
The diffusion term in equation (\ref{skilling}), $\boldsymbol{\nabla \cdot} ( \kappa_{\text{p}}  \boldsymbol{n}\boldsymbol{n} \boldsymbol{\cdot \nabla} f_{p})$, shows that the diffusion follows the magnetic field lines. 

From equation (\ref{skilling}), one can easily derive
\begin{equation}
\frac{\partial f_{p}}{\partial t} +\boldsymbol{\nabla \cdot} \boldsymbol{S}-Q+
\frac{1}{3 p^{2}}\frac{\partial}{\partial p}[p^{3}(\boldsymbol{u}+\boldsymbol{v_{\text{A}}})\boldsymbol{\cdot \nabla} f_{p}]=0 \label{distcons} \text{,}
\end{equation}
where we have defined the particle current \citep[e.g.][]{1983SSRv...36....3V, 1979Ap&SS..60..335W}
\begin{equation}
\boldsymbol{S}=(\boldsymbol{u}+\boldsymbol{v}_{\text{A}})f_{p}- \kappa_{\text{p}}  \boldsymbol{n}\boldsymbol{n} \boldsymbol{\cdot \nabla} f_{p} -\frac{1}{3p^{2}}(\boldsymbol{u}+\boldsymbol{v}_{\text{A}})\frac{\partial }{\partial p}(p^{3}f_{p})\text{.}
\label{current}
\end{equation}
Integration of equation (\ref{distcons}) over all particle momenta results in the equation for the number density of cosmic rays (defined as $n_{\text{CR}}=\int_{0}^{\infty} f_{p}4\pi p^{2}dp$)
\begin{equation}
\frac{\partial n_{\text{CR}}}{\partial t}+\boldsymbol{\nabla \cdot}\left(
\int_{0}^{\infty} 4\pi p^{2}\boldsymbol{S}dp\right)-\int_{0}^{\infty}4\pi p^{2}Qdp=0\text{.}
\end{equation}

To derive the energy equation for cosmic rays, we define three macroscopic quantities: the cosmic-ray pressure $P_{\text{c}}$, the cosmic-ray energy density $E_{\text{c}}$ and the cosmic-ray energy flux $\boldsymbol{F}_{\text{c}}$ as the following moments of $f_{p}$ and $\boldsymbol{S}$:
\begin{align}
P_{\text{c}}&=\frac{4\pi}{3}\int_{0}^{\infty}vp^{3}f_{p}dp   \label{defin1}\\
E_{\text{c}}&=4\pi\int_{0}^{\infty}p^{2}T_{\text{p}}f_{p}dp   \label{defin2}\\
\boldsymbol{F}_{\text{c}}&=4\pi\int_{0}^{\infty}p^{2}T_{\text{p}}\boldsymbol{S}dp    \label{defin3} \text{,}
\end{align}
where $T_{\text{p}}(p)$ is the kinetic energy of a cosmic-ray particle with momentum $p$ and mass $m$,
\begin{equation}
T_{\text{p}}(p)=\biggl[\left(1+\frac{p^{2}}{m^{2}c^{2}}\right)^{1/2}-1\biggr]mc^{2}  \text{.}
\end{equation}

By taking the appropriate moments of equations (\ref{skilling}) and (\ref{current}) and using the definitions in equations (\ref{defin1}), (\ref{defin2}) and (\ref{defin3}), we obtain 
\begin{align}
\boldsymbol{F}_{\text{c}}&=(E_{\text{c}}+P_{\text{c}})(\boldsymbol{u}+\boldsymbol{v}_{\text{A}})-\boldsymbol{n}\kappa_{\text{c}}(\boldsymbol{n \cdot \nabla} E_{\text{c}}) \label{energyflux}\text{,} \\
\frac{\partial E_{\text{c}}}{\partial t}&=(\boldsymbol{u}+\boldsymbol{v}_{\text{A}})\boldsymbol{\cdot \nabla} P_{\text{c}}-\boldsymbol{\nabla \cdot F}_{\text{c}} + \bar{Q}   \text{,}  \label{firstenergy}
\end{align}
where $\kappa_{\text{c}}(\boldsymbol{x})$ represents an effective diffusion coefficient
\begin{equation}
\kappa_{\text{c}}(\boldsymbol{x})=\frac{\int_{0}^{\infty}p^{2}T_{\text{p}}\kappa_{\text{p}} (\boldsymbol{x},p)(\boldsymbol{n \cdot \nabla} f_{p})dp} { \int_{0}^{\infty}p^{2}T_{\text{p}}(\boldsymbol{n \cdot \nabla} f_{p}) dp} \text{,}
\end{equation}
and $\bar{Q}$ is the net source of mean kinetic energy density of cosmic rays
\begin{equation}
\bar{Q}=4\pi\int_{0}^{\infty}p^{2}T_{\text{p}}Qdp\text{.}
\end{equation}
The fist term in the right-hand side of equation (\ref{firstenergy}) represents the energy-loss rate of cosmic rays due to the work of cosmic-ray pressure gradient on the background plasma ($\boldsymbol{u \cdot \nabla} P_{\text{c}}$) and the generation of Alfv$\acute{\text{e}}$n waves ($\boldsymbol{v}_{\text{A}}\boldsymbol{\cdot \nabla} P_{\text{c}}$, see \citet{1982A&A...116..191M}). 

Assuming that cosmic rays are ultra-relativistic ($v\approx c$), from equations (\ref{defin1}) and (\ref{defin2}), we find $P_{\text{c}}=(\gamma_{\text{c}}-1)E_{\text{c}}$, where $\gamma_{\text{c}}=4/3$ is the adiabatic index for the cosmic rays ($\gamma_{\text{c}}=5/3$ for non-relativistic particles). From equations (\ref{energyflux}) and (\ref{firstenergy}), we thus obtain the time-dependent cosmic-ray equations 
\begin{align}
\boldsymbol{F}_{\text{c}}&=\gamma_{\text{c}}E_{\text{c}}(\boldsymbol{u}+\boldsymbol{v}_{\text{A}})- \boldsymbol{n}\kappa_{\text{c}}(\boldsymbol{n \cdot \nabla} E_{\text{c}}) \label{energyflux2} \text{,}\\
\frac{\partial E_{\text{c}}}{\partial t} &=(\gamma_{\text{c}}-1) 
(\boldsymbol{u}+\boldsymbol{v}_{\text{A}})\boldsymbol{\cdot \nabla} E_{\text{c}}
- \boldsymbol{\nabla \cdot F}_{\text{c}} + \bar{Q}    \label{firstenergy2} \text{.}
\end{align}

%\bsp 
\label{lastpage}

\end{document}